\documentclass[english]{article}
\usepackage[T1]{fontenc}
\usepackage[latin9]{inputenc}
\usepackage{geometry}
\usepackage{color}
\usepackage{babel}
\usepackage{array}
\usepackage{float}
\usepackage{url}
\usepackage{bm}
\usepackage{multirow}
\usepackage{amsmath}
\usepackage{amssymb}
\usepackage{graphicx}
\usepackage{setspace}
\usepackage[unicode=true,pdfusetitle,
 bookmarks=true,bookmarksnumbered=false,bookmarksopen=false,
 breaklinks=false,pdfborder={0 0 0},pdfborderstyle={},backref=false,colorlinks=false]
 {hyperref}

\makeatletter

\providecommand{\tabularnewline}{\\}
\floatstyle{ruled}
\newfloat{algorithm}{tbp}{loa}
\providecommand{\algorithmname}{Algorithm}
\floatname{algorithm}{\protect\algorithmname}

\newenvironment{lyxlist}[1]
	{\begin{list}{}
		{\settowidth{\labelwidth}{#1}
		 \setlength{\leftmargin}{\labelwidth}
		 \addtolength{\leftmargin}{\labelsep}
		 }}
	{\end{list}}

\usepackage{algorithm,algpseudocode}

\usepackage{animate}

\usepackage{xcolor}

\algnewcommand\algorithmicforeach{\textbf{for each}}
\algdef{S}[FOR]{ForEach}[1]{\algorithmicforeach\ #1\ \algorithmicdo}
\algdef{SE}[SUBALG]{Indent}{EndIndent}{}{\algorithmicend\ }%
\algtext*{Indent}
\algtext*{EndIndent}

\@ifundefined{showcaptionsetup}{}{%
 \PassOptionsToPackage{caption=false}{subfig}}
\usepackage{subfig}
\makeatother

\begin{document}
\title{Efficient smoothed particle radiation hydrodynamics I:\\
 Thermal radiative transfer\vspace{0.1\textheight}
}
\author{Brody R. Bassett$^{a,\star}$, J. Michael Owen$^{a}$, Thomas A. Brunner$^{a}$\\
\\
{\normalsize{}\href{mailto:bassett4@llnl.gov}{bassett4@llnl.gov},
\href{mailto:owen8@llnl.gov}{owen8@llnl.gov}, \href{mailto:brunner6@llnl.gov}{brunner6@llnl.gov}}\\
{\normalsize{}}\\
{\normalsize{}$^{a}$Lawrence Livermore National Laboratory}\\
{\normalsize{}7000 East Avenue, Livermore, CA, 94550}\\
{\normalsize{}}\\
{\normalsize{}$^{\star}$Corresponding author}\vspace{0.1\textheight}
}
\date{$ $}
\maketitle
\begin{abstract}
This work presents efficient solution techniques for radiative transfer
in the smoothed particle hydrodynamics discretization. Two choices
that impact efficiency are how the material and radiation energy are
coupled, which determines the number of iterations needed to converge
the emission source, and how the radiation diffusion equation is solved,
which must be done in each iteration. The coupled material and radiation
energy equations are solved using an inexact Newton iteration scheme
based on nonlinear elimination, which reduces the number of Newton
iterations needed to converge within each time step. During each Newton
iteration, the radiation diffusion equation is solved using Krylov
iterative methods with a multigrid preconditioner, which abstracts
and optimizes much of the communication when running in parallel.
The code is verified for an infinite medium problem, a one-dimensional
Marshak wave, and a two and three-dimensional manufactured problem,
and exhibits first-order convergence in time and second-order convergence
in space. For these problems, the number of iterations needed to converge
the inexact Newton scheme and the diffusion equation is independent
of the number of spatial points and the number of processors. \\
\end{abstract}

\pagebreak{}

\section{Introduction}

Smoothed particle hydrodynamics (SPH) is a meshless approach to solving
the hydrodynamics equations, in which the fluid is separated into
discrete masses that are used as interpolation points (for an overview,
see Refs. \cite{monaghan2005smoothed,liu2010smoothed}). SPH has several
\textcolor{black}{desirable} properties, such as automatic conservation
of mass and enforced conservation of energy and momentum, flexible
point topology, and Galilean invariance. Like mesh-based Lagrangian
codes, the resolution of the problem follows the mass, but unlike
mesh-based codes, SPH does not have issues with mesh tangling. Drawbacks
of SPH include a lack of zeroth-order consistency (the interpolant
cannot generally reproduce a constant exactly) \cite{morris1996study}
and additional expense that comes from computing connectivity at each
time step. There are extensions to SPH that correct some of these
issues, such as moving least squares particle hydrodynamics \cite{dilts1999moving,dilts2000moving}
and conservative reproducing kernel smoothed particle hydrodynamics
\cite{frontiere2017crksph}. 

The thermal radiative transfer equations have been solved using a
smoothed particle hydrodynamics discretization previously. The most
popular method is implicit, two-temperature, flux-limited diffusion,
first implemented in Refs. \cite{whitehouse2004smoothed,whitehouse2005faster}.
These papers assume an ideal gas equation of state. In the latter
paper, the system is reduced to solving a quartic equation per point
within a Gauss-Seidel iteration scheme. \textcolor{black}{This method has been extended to include the physics of a diffuse
interstellar medium for star formation \cite{bate2015combining}.} There are other implementations of flux-limited diffusion, most of
which have been applied to astrophysical problems \cite{viau2006implicit,mayer2007fragmentation}.
The form of the second derivative used most often for the radiation
diffusion second derivative was first applied to heat diffusion \cite{brookshaw1985method}.
The optically-thin variable Eddington factor equations with Eddington
factors determined by source information (as opposed to by a full
transport calculation) have been applied to cosmological simulation
\cite{petkova2009implementation}. Similar methods that have been
studied include ray-tracing \cite{altay2008sphray}, Monte Carlo \cite{nayakshin2009dynamic},
and neutrino flux-limited diffusion \cite{fryer2006snsph}. 

Thermal radiative transfer couples the transport of photons with the
hydrodynamic state. Common coupling techniques for the radiation and
material equations include simple convergence of the residuals of
both equations (including Newton-Krylov methods), using a single Newton
iteration, lagging the nonlinear terms in the equation, predictor-corrector
schemes, and linearization of the nonlinear emission source \cite{lowrie2004comparison,mousseau2000physics}.
\textcolor{black}{Another common method is linear multifrequency-grey acceleration,
which accelerates multigroup convergence and can be used as a preconditioner
for a Krylov iterative solve \cite{morel1985synthetic,morel2007linear}.
For more information on time discretization methods, see the following
papers, which compare time integration methods for radiative transfer:
block Jacobi, Schur complement, and operator splitting approaches
\cite{brown2001preconditioning}; Newton's method, Newton-Krylov,
and linearized approaches \cite{lowrie2004comparison}; and Newton's
method, linearized approaches, and operator splitting \cite{tetsu2016comparison}.
A common thread among these is that while full Newton's method is
expensive, it is more stable than the alternatives. The method used
in this paper is based on nonlinear elimination methods, as described
in Ref. \cite{lanzkron1996analysis} and applied to radiation diffusion
in Ref. \cite{brunner2020nonlinear}, which fully converges the nonlinear
solution with only a marginal cost increase over the linearized solution. }

The radiation diffusion and transport equations have also been solved
using meshless methods other than SPH, including for coupled radiative
transport and conductive heat transfer \cite{sadat2006use,liu2007meshless,kindelan2010application},
neutron transport \cite{bassett2019meshless}, and neutron diffusion
\cite{rokrok2012element,tanbay2013numerical}. Many of these discretizations
involve either relatively flat meshless functions (which increases
accuracy but makes the system ill-conditioned) or integration of the
meshless functions. The advantage of using the SPH discretization
directly is that the same functions that are used for hydrodynamics
can be reused for the radiation, without recomputing the topology. 

\textcolor{black}{Some particle methods map unknowns to a background grid for certain
physics, such as electromagnetic particle-in-cell methods that map
the charge of the particles to a background grid for the solution
of Maxwell's equations \cite{birdsall1991particle}. A similar method
could be employed for radiation diffusion with SPH, in which the density
and material energy would be mapped to a background mesh for a radiation
diffusion solve, but this would introduce complications with mesh
creation and diffusivity due to the mapping. For consistent SPH diffusion,
the solution points are the same for hydrodynamics and radiation,
which simplifies the solution process and eliminates possible errors
due to mapping. SPH diffusion is not computationally competitive with
mesh-based diffusion due to the large number of neighbors, but is
compatible with SPH hydrodynamics, which can handle problems that
mesh-based methods struggle with. }

The goal of this research is to make SPH radiation diffusion more
efficient. This is done by improving the solution of the diffusion
problem using fast and accurate preconditioners and applying material-radiation
coupling methods that speed up the convergence of the radiation emission
and absorption. With appropriate time step constraints, the solver
is stable and exhibits first-order convergence in time and second-order
convergence in space to analytic solutions and manufactured problems.
The methods used here scale well with the number of points and under
domain decomposition due to the use of fast multigrid solvers \cite{falgout2002hypre},
which also makes the method simple to implement on distributed architectures.
The remainder of this paper is structured as follows. In Sec. \ref{sec:theory},
the SPH thermal radiative transfer equations are derived and discretized
in time and space. In Sec. \ref{sec:methodology}, the implementation
of the equations is discussed. Finally, in Sec. \ref{sec:results},
results for problems in one, two and three dimensions with known solutions
are presented to verify the accuracy of the code. In Paper II, the
full SPH radiation hydrodynamics equations are derived with appropriate
radiation-material coupling terms, and results for full radiation
hydrodynamics are presented. 

\section{Theory\label{sec:theory}}

The smoothed particle hydrodynamics method is used to discretize the
thermal radiative transfer equations in space. The discretization
in time reduces a coupled set of equations for the material and radiation
energies to separate updates for the material and radiation energies
in an iterative process. The resulting equations conserve energy. 

\subsection{The thermal radiative transfer equations}

The thermal radiative transfer equations are \begin{subequations}
\begin{gather}
\rho\partial_{t}e=-c\sigma_{a}B+c\sigma_{a}E+Q_{e},\label{eq:material-energy}\\
\partial_{t}E=-\partial_{x}^{\alpha}F^{\alpha}-c\sigma_{a}E+c\sigma_{a}B+Q_{E},\label{eq:zeroth-moment}\\
\frac{1}{c}\partial_{t}F^{\alpha}=-c\partial_{x}^{\beta}P^{\alpha\beta}-\sigma_{t}F^{\alpha},\label{eq:first-moment}
\end{gather}
\end{subequations}with the variables
\begin{lyxlist}{00.00.0000}
\item [{$t$,}] time,
\item [{$x$,}] position,
\item [{$\rho$,}] mass density,
\item [{$e$}] specific material energy,
\begin{spacing}{0.7}
\item [{$E$,}] radiation energy density,
\end{spacing}
\item [{$F^{\alpha}$,}] radiation flux, 
\item [{$P^{\alpha\beta}$,}] radiation pressure,
\begin{spacing}{0.7}
\item [{$T$,}] material temperature,
\end{spacing}
\item [{$B$,}] integrated photon emission function,
\item [{$c$,}] speed of light in a vacuum,
\begin{spacing}{0.7}
\item [{$\sigma_{t}$,}] total opacity,
\item [{$\sigma_{a}$,}] absorption opacity,
\item [{$a$,}] black-body constant,
\item [{$Q_{e}$,}] nonhomogeneous material energy source,
\item [{$Q_{E}$,}] nonhomogeneous radiation energy source. 
\end{spacing}
\end{lyxlist}
\textcolor{black}{Greek letters used as superscripts (e.g. $F^{\alpha}$) indicate dimensional
components of a vector. For derivatives, a similar definition holds,
where $\partial_{x}^{\alpha}E$ would indicate the $\alpha$ component
of the gradient of $E$ with respect to $x$. Repeated indices indicate
summation, so $\partial_{x}^{\alpha}F^{\alpha}$ is the divergence
of $F$ with respect to $x$. }

The radiation transport equation has been integrated over all energy
frequencies (the grey approximation) and integrated over angle to
produce the first two angular moments. For the derivation in Paper
I, the material is assumed to be stationary. The emission term is
defined as 
\begin{equation}
B=aT^{4},
\end{equation}
where the black body constant $a$ is defined in terms of the Stefan-Boltzmann
constant $\sigma_{SB}$ and the speed of light $c$ as $a=4\sigma_{SB}/c$.
\textcolor{black}{The material energy and temperature are connected through an equation
of state. An example is the ideal gas equation of state, in which
the temperature is proportional to the energy, 
\begin{equation}
T=\frac{\mu\left(\gamma-1\right)m_{p}}{k_{B}}e,\label{eq:ideal-eos}
\end{equation}
where $\mu$ is the molecular mass, $k_{B}$ is the Boltzmann constant,
and $\gamma$ is the ratio of heat capacities. The opacities generally
depend on material, temperature and density. }

Assuming the specific intensity is linearly anisotropic in angle,
the pressure term becomes 
\begin{equation}
P^{\alpha\beta}=\frac{1}{3}I_{\alpha\beta}E,
\end{equation}
where $I$ is the tensor identity. Assuming that the time derivatives
are neglected, the first angular moment of the transport equation
{[}Eq. (\ref{eq:zeroth-moment}){]} can be solved for the radiation
flux to get Fick's law,
\begin{equation}
F^{\alpha}=-\frac{c}{3\sigma_{t}}\partial_{x}^{\alpha}E.\label{eq:ficks}
\end{equation}
Replacing the flux term in the zeroth angular moment equation {[}Eq.
(\ref{eq:first-moment}){]} with Fick's law and replacing the pressure
term by the linearly anisotropic value results in the diffusion equation,
\begin{equation}
\partial_{t}E=\partial_{x}^{\alpha}\frac{c}{3\sigma_{t}}\partial_{x}^{\alpha}E-c\sigma_{a}E+c\sigma_{a}B+Q_{E}.
\end{equation}

In the original radiation equations {[}Eqs. (\ref{eq:zeroth-moment})
and (\ref{eq:first-moment}){]}, if the pressure were not isotropic,
the radiation would propagate correctly at the speed of light. However,
when the time derivative on the first moment equation is dropped,
this adds an error that allows the radiation to propagate faster than
the speed of light, $\left|\bm{F}\right|>cE$. This effect can be
prevented by applying a flux limiter $\lambda$ to the diffusion equation,
\begin{equation}
\partial_{t}E=\partial_{x}^{\alpha}\frac{c\lambda}{\sigma_{t}}\partial_{x}^{\alpha}E-c\sigma_{a}E+c\sigma_{a}B+Q_{E}.\label{eq:diffusion}
\end{equation}
With a constant $\lambda=1/3$, the original equation is recovered.
The flux limiter is usually written in terms of the constant $R$,
\begin{equation}
R=\frac{\sqrt{\left(\partial_{x}^{\alpha}E\right)\left(\partial_{x}^{\alpha}E\right)}}{\sigma_{t}E},
\end{equation}
with examples such as the Levermore flux limiter \cite{levermore1984relating},
\begin{equation}
\lambda=\frac{2+R}{6+3R+R^{2}},
\end{equation}
and the Larsen flux limiter \cite{morel2000diffusion},
\begin{equation}
\lambda=\frac{1}{\left(3^{k}+R^{k}\right)^{1/k}},
\end{equation}
where $k$ is a chosen constant. Both of these have the correct limits
for diffusive regions $\left(R\to0\right)$, which is the diffusion
equation, and optically thin regions $\left(R\to\infty\right)$, which
describes radiation propagating at the speed of light \cite{mihalas2013foundations},
\begin{equation}
F^{\alpha}\to\begin{cases}
-\frac{c}{3\sigma_{t}}\partial_{x}^{\alpha}E, & R\to0,\\
-\frac{\partial_{x}^{\alpha}E}{\sqrt{\left(\partial_{x}^{\beta}E\right)\left(\partial_{x}^{\beta}E\right)}}cE, & R\to\infty.
\end{cases}
\end{equation}
The Larsen flux limiter with $k=2$ is used in most of the results
in this paper.

\subsection{Time discretization}

The coupled material and radiation energy equations {[}Eqs. (\ref{eq:material-energy})
and (\ref{eq:diffusion}){]} are discretized fully implicitly in time,
using the nonlinear elimination methodology described in Ref. \cite{brunner2020nonlinear}.
The derivation begins with the coupled material and radiation diffusion
equations discretized using backward Euler,
\begin{equation}
\bm{G}=\left[\begin{array}{c}
m\left(e^{n},E^{n}\right)\\
r\left(e^{n},E^{n}\right)
\end{array}\right]=\left[\begin{array}{c}
\dfrac{\rho}{\Delta t}\left(e^{n}-e^{n-1}\right)+c\sigma_{a}B^{n}-c\sigma_{a}E^{n}-Q_{e}^{n}\\
\dfrac{1}{\Delta t}\left(E^{n}-E^{n-1}\right)-\partial_{x}^{\alpha}\dfrac{c\lambda}{\sigma_{t}}\partial_{x}^{\alpha}E^{n}+c\sigma_{a}E^{n}-c\sigma_{a}B^{n}-Q_{E}^{n}
\end{array}\right]=\left[\begin{array}{c}
0\\
0
\end{array}\right],\label{eq:ne-starting-point}
\end{equation}
where $\bm{U}=\left[\begin{array}{cc}
e & E\end{array}\right]^{\top}$. Newton's method updates the state according to\begin{subequations}\label{eq:regular-newton}
\begin{gather}
\bm{J}^{\ell}\bm{\delta}^{\ell}=-\bm{G}^{\ell},\\
\bm{U}^{\ell+1}=\bm{U}^{\ell}+\bm{\delta}^{\ell},\label{eq:newton-update}
\end{gather}
 where $\bm{J}$ is the Jacobian with respect to $e$ and $E$, 
\begin{equation}
\bm{J}=\left[\begin{array}{cc}
J_{m,e} & J_{m,E}\\
J_{r,e} & J_{r,E}
\end{array}\right]=\left[\begin{array}{cc}
\dfrac{\rho}{\Delta t}+\dfrac{c}{c_{v}}\sigma_{a}\partial_{T}B & -c\sigma_{a}\\
-\dfrac{c}{c_{v}}\sigma_{a}\partial_{T}B & \dfrac{1}{\Delta t}-\partial_{x}^{\alpha}\frac{c\lambda}{\sigma_{t}}\partial_{x}^{\alpha}+c\sigma_{a}
\end{array}\right],
\end{equation}
\end{subequations}and $\bm{\delta}=\left[\delta_{e},\delta_{E}\right]^{\top}$
is the change in the state variables and $c_{v}=\frac{\partial e}{\partial T}$
is the specific heat capacity. 

To simplify the Newton iteration, the Schur compliment of the Jacobian
is taken, resulting in the Jacobian system 
\begin{equation}
\left[\begin{array}{cc}
J_{m,e} & J_{m,E}\\
0 & J_{r,E}-J_{r,e}J_{m,e}^{-1}J_{m,E}
\end{array}\right]\left[\begin{array}{c}
\delta_{e}\\
\delta_{E}
\end{array}\right]=\left[\begin{array}{c}
-G_{m}\\
-G_{r}+J_{r,e}J_{m,e}^{-1}G_{m}
\end{array}\right]\label{eq:newton-schur}
\end{equation}
that can be solved first for $\delta_{E}$ and then $\delta_{e}$.
This system of equations can be first solved for $\delta_{E}$ and
then $\delta_{e}$. The solution in Eq. (\ref{eq:newton-schur}) combined
with the update in Eq. (\ref{eq:newton-update}) is full Newton iteration
without approximation. Noting that $\partial_{T}B=4aT^{3}$, the term
$J_{r,e}J_{m,e}^{-1}$ can be written in terms of the Fleck factor
commonly used in implicit Monte Carlo \cite{fleckjr1971implicit},
\begin{equation}
f=\left(1+c\sigma_{a}\Delta t\frac{4aT^{3}}{\rho c_{v}}\right)^{-1},
\end{equation}
which is used to forward-predict the emission source, as 
\begin{equation}
J_{r,e}J_{e,e}^{-1}=f-1.
\end{equation}
Finally, the temperature is held constant in $f$ during each time
step, which results in an inexact Newton iteration scheme,
\begin{equation}
\dfrac{1}{\Delta t}E^{n,\ell+1}-\partial_{x}^{\alpha}\dfrac{c\lambda}{\sigma_{t}}\partial_{x}^{\alpha}E^{n,\ell+1}+c\sigma_{a}fE^{n,\ell+1}=\dfrac{1}{\Delta t}E^{n-1}+c\sigma_{a}B^{n,\ell}-\left(1-f\right)c\sigma_{a}E^{n,\ell}+Q_{E}^{n},
\end{equation}
with the time step index $n$ and the iteration index $\ell$. When
$E^{n}$ has converged, the terms involving $f$ in the diffusion
equation cancel out and the original, time-discretized diffusion equation
is recovered.

Equations (\ref{eq:regular-newton}) represent the standard Newton
solution method for the thermal radiative transfer equations. Here,
a nonlinear acceleration technique from Ref. \cite{brunner2020nonlinear}
is additionally applied. The following two functions are helpful in
describing the iterative process of solving the equations,\begin{subequations}\label{eq:ne-equations}
\begin{gather}
m\left(e^{\star},E\right)=\dfrac{\rho}{\Delta t}\left(e^{\star}-e^{n-1}\right)+c\sigma_{a}B^{\star}-c\sigma_{a}E-Q_{e}^{n}=0,\\
r^{\dagger}\left(e,E,E^{\star}\right)=\dfrac{1}{\Delta t}E^{\star}-\partial_{x}^{\alpha}\dfrac{c\lambda}{\sigma_{t}}\partial_{x}^{\alpha}E^{\star}+c\sigma_{a}fE^{\star}-\dfrac{1}{\Delta t}E^{n-1}-c\sigma_{a}B+\left(1-f\right)c\sigma_{a}E-Q_{E}^{n}=0
\end{gather}
\end{subequations}where $e^{\star}$ and $E^{\star}$ represent the
values to be solved for. Note the use of $r^{\dagger}$ to distinguish
the diffusion equation from the starting equation $r$. The solution
of $r^{\dagger}\left(e,E,E^{\star}\right)=0$ represents solving a
spatially-coupled linear equation, as discussed in Sec. \ref{subsec:mat-rad-solve}.
Note that the value of the material energy in these equations is entirely
dependent on the radiation energy, which means that a solution of
$r^{\dagger}$ represents an inexact Newton iteration of the full
material and radiation energy system. After a solution of $r^{\dagger}$
for the radiation energy, the material energy is updated by solving
$m\left(e^{*},E\right)=0$, which represents a nonlinear solve using
Newton's method,
\begin{equation}
e^{\star}\leftarrow e^{\star}-\left(J_{m,e}^{\star}\right)^{-1}\left[\dfrac{\rho}{\Delta t}\left(e^{\star}-e^{n-1}\right)+c\sigma_{a}B^{\star}-c\sigma_{a}E-Q_{e}^{n}\right],\label{eq:full-material-solve}
\end{equation}
where $\leftarrow$ represents an update to the value of $e^{\star}$.
This update is performed until $e^{\star}$ is converged. This differs
from standard Newton iteration, where the material energy update would
be done simultaneously with the radiation energy update and without
solving $m\left(e^{*},E\right)=0$ until $e^{*}$ is converged. The
iterative solution procedure for these two equations is described
in Alg. \ref{alg:mat-rad-solve}. \textcolor{black}{Note that because the material energy has been eliminated through
the nonlinear elimination process and can be considered a function
of the radiation energy $E$, not an independent variable. This means
that when the radiation energy is updated, the material energy is
recomputed. The initial update makes the material energy consistent
with the value of the radiation energy before iterations start. }

The outer iterations, or the inexact Newton iteration of the material
and radiation energy system, continue until the convergence criteria
are met,\begin{subequations}\label{eq:mat-rad-convergence}
\begin{gather}
\left|\frac{e^{n,\ell+1}-e^{n,\ell}}{e^{n,\ell+1}}\right|<\epsilon_{e},\\
\left|\frac{E^{n,\ell+1}-E^{n,\ell}}{E^{n,\ell+1}}\right|<\epsilon_{E}.
\end{gather}
\end{subequations}The inner iterations, or the iterations required
to converge $m$ and $r$ within an outer iteration, use similar equations
for determining convergence. As the absorption term appears with both
$E^{n,\ell}$ and $E^{n,\ell+1}$, convergence is required for energy
conservation to be preserved (see Sec. \ref{subsec:conservation}). 

\subsection{Introduction to smoothed particle hydrodynamics\label{subsec:sph-intro}}

\textcolor{black}{In this section, the standard SPH derivatives needed for the thermal
radiative transfer approximation are derived. For a more complete
review of the SPH approximation, see Ref. \cite{monaghan2005smoothed}.} 

Smoothed particle hydrodynamics involves interpolation of fields using
kernels, which are functions centered at interpolation points. The
initial assumption needed is that SPH kernels approximate delta functions,
or $W\left(x-x^{\prime},h\right)\to\delta\left(x-x^{\prime}\right)$
as $h\to0$, where $h$ is the smoothing parameter that determines
the width of the kernels,
\begin{align}
g\left(x\right) & =\int_{V}\delta\left(x-x^{\prime}\right)g\left(x^{\prime}\right)dV'\nonumber \\
 & \approx\int_{V}W\left(x-x^{\prime},h\right)g\left(x^{\prime}\right)dV'.
\end{align}
This also means that the integral of the kernel should be equal to
one,
\begin{equation}
\int_{V}W\left(x-x^{\prime},h\right)dV'=1.\label{eq:kernel-normalization}
\end{equation}

In practice, the $h$ can depend on both $x$ and $x^{\prime}$. The
second assumption is that a set of kernels in space can be used to
form a quadrature, with abscissas at the kernel centers $\bm{x}_{i}$
and weights equal to the kernel volume $V_{i}$. With these two assumptions,
the kernels can be used to interpolate a function,

\begin{align}
\left\langle g\left(x\right)\right\rangle  & =\int_{V}W\left(x-x^{\prime},h\right)g\left(x^{\prime}\right)dV'\nonumber \\
 & \approx\sum_{j}V_{j}W\left(x-x_{j},h\right)g_{j}.\label{eq:sph-interpolant}
\end{align}
Here $g_{j}$ is the function evaluated at point $j$, $g_{j}=g\left(x_{j}\right)$,
and $\left\langle \cdot\right\rangle $ indicates an interpolated
quantity. The interpolant is used to calculate derivatives of the
field. This interpolant does not have the Kronecker delta property,
\begin{equation}
\left\langle g\left(x\right)\right\rangle _{i}=\sum_{j}V_{j}W\left(x_{i}-x_{j},h\right)g_{j}\neq g_{i}.
\end{equation}

To approximate a derivative in SPH, another three properties are needed
from the kernel: that the kernel does not intersect a boundary, that
the derivative of the kernel is antisymmetric in $\bm{x}$ and $\bm{x}'$,
and that the integral of the derivative of the kernel is zero. With
these approximations, the derivative of a function can be approximated
as 

\begin{align}
\left\langle \partial_{x}^{\alpha}g\left(x\right)\right\rangle  & =\int_{V}W\left(x-x^{\prime},h\right)\partial_{x^{\prime}}^{\alpha}g\left(x^{\prime}\right)dV'\nonumber \\
 & =-\int_{V}\partial_{x^{\prime}}^{\alpha}W\left(x-x^{\prime},h\right)g\left(x^{\prime}\right)dV'\nonumber \\
 & =\partial_{x}^{\alpha}\int_{V}W\left(x-x^{\prime},h\right)g\left(x^{\prime}\right)dV'-g\left(x\right)\partial_{x}^{\alpha}\int_{V}W\left(x-x^{\prime},h\right)dV'\nonumber \\
 & \approx\sum_{j}V_{j}\left(g_{j}-g\left(x\right)\right)\partial_{x}^{\alpha}W\left(x-x_{j},h\right).\label{eq:sph-derivative}
\end{align}
Note that because of the added $g\left(x\right)$ term (which equals
zero in integral form due to the derivative), this approximation of
the derivative goes to zero for a constant function. 

For a diffusion-like problem, a second-order series expansion for
$g\left(x\right)$ about $x^{\prime}$ and $g\left(x^{\prime}\right)$
about $x$,
\begin{equation}
\partial_{x^{\prime}}^{\alpha}g\left(x^{\prime}\right)\approx\frac{x^{\alpha}-x^{\alpha,\prime}}{\left(x^{\beta}-x^{\beta,\prime}\right)\left(x^{\beta}-x^{\beta,\prime}\right)}\left[g\left(x^{\prime}\right)-g\left(x\right)\right],
\end{equation}
can be used to produce a commonly-used second-order approximation
to the second derivative \cite{monaghan2005smoothed},
\begin{align}
\left\langle \partial_{x}^{\alpha}\left[k\left(x\right)\partial_{x}^{\alpha}g\left(x\right)\right]\right\rangle  & =\int_{V}W\left(x-x^{\prime},h\right)\partial_{x^{\prime}}^{\alpha}\left[k\left(x^{\prime}\right)\partial_{x^{\prime}}^{\alpha}g\left(x^{\prime}\right)\right]dV'\nonumber \\
 & =-\int_{V}k\left(x^{\prime}\right)\partial_{x^{\prime}}^{\alpha}W\left(x-x^{\prime},h\right)\partial_{x^{\prime}}^{\alpha}g\left(x^{\prime}\right)dV'\nonumber \\
 & =\partial_{x}^{\alpha}\int_{V}k\left(x^{\prime}\right)W\left(x-x^{\prime},h\right)\partial_{x^{\prime}}^{\alpha}g\left(x^{\prime}\right)dV'+k\left(x\right)\partial_{x}^{\alpha}g\left(x\right)\partial_{x}^{\alpha}\int_{V}W\left(x-x^{\prime},h\right)dV'\nonumber \\
 & =\int_{V}\left[k\left(x\right)\partial_{x}^{\alpha}g\left(x\right)+k\left(x^{\prime}\right)\partial_{x^{\prime}}^{\alpha}g\left(x^{\prime}\right)\right]\partial_{x}^{\alpha}W\left(x-x^{\prime},h\right)dV'\nonumber \\
 & \approx\int_{V}\left[k\left(x\right)+k\left(x^{\prime}\right)\right]\left[g\left(x\right)-g\left(x^{\prime}\right)\right]\frac{x^{\alpha}-x^{\alpha,\prime}}{\left(x^{\beta}-x^{\beta,\prime}\right)\left(x^{\beta}-x^{\beta,\prime}\right)}\partial_{x}^{\alpha}W\left(x-x^{\prime},h\right)dV'\nonumber \\
 & \approx\sum_{j}V_{j}\left(k\left(x\right)+k_{j}\right)\left(g\left(x\right)-g_{j}\right)\frac{x^{\alpha}-x_{j}^{\alpha}}{\left(x^{\beta}-x_{j}^{\beta}\right)\left(x^{\beta}-x_{j}^{\beta}\right)}\partial_{x}^{\alpha}W\left(x-x_{j},h\right).\label{eq:sph-second-derivative}
\end{align}
Like the first derivative approximation, the second derivative is
zero for a constant function because the $g$ terms cancel. 

When discretizing an equation in SPH, the equation is first multiplied
by the kernel $W\left(x-x^{\prime},h\right)$ and integrated, similar
to the finite element method. The distinction is that the interpolant
is only used to calculate derivatives. The terms without derivatives
are approximated using the definition of the interpolant, 
\begin{equation}
\left\langle g\left(x\right)\right\rangle =\int_{V}W\left(x-x^{\prime},h\right)g\left(x^{\prime}\right)dV'\approx g\left(x\right).\label{eq:interpolant-approx}
\end{equation}
The terms with derivatives are approximated as derivatives of the
interpolant, as in Eq. \ref{eq:sph-derivative}. The final step in
an SPH discretization is to add constraints to match the free variables
in the equations. This can be done by multiplying the equations by
a delta function $\delta\left(x-x_{i}\right)$ and integrating, or
equivalently, evaluating the functions at $x=x_{i}$. This results
in the fully-discrete approximations to Eqs. (\ref{eq:sph-interpolant}),
(\ref{eq:sph-derivative}), and (\ref{eq:sph-second-derivative}),\begin{subequations}
\begin{gather}
\left\langle g\left(x\right)\right\rangle _{i}\approx\sum_{j}V_{j}W_{ij}g_{j},\\
\left\langle \partial_{x}^{\alpha}g\left(x\right)\right\rangle _{i}\approx\sum_{j}V_{j}\left(g_{j}-g_{i}\right)\partial_{x_{i}}^{\alpha}W_{ij},\\
\left\langle \partial_{x}^{\alpha}\left[k\left(x\right)\partial_{x}^{\alpha}g\left(x\right)\right]\right\rangle _{i}\approx\sum_{j}V_{j}\left(k_{i}+k_{j}\right)\left(g_{i}-g_{j}\right)\frac{x_{ij}^{\alpha}}{x_{ij}^{\beta}x_{ij}^{\beta}}\partial_{x_{i}}^{\alpha}W_{ij},
\end{gather}
\end{subequations}where \begin{subequations}
\begin{gather}
x_{ij}^{\alpha}=x_{i}^{\alpha}-x_{j}^{\alpha},\\
W_{ij}=W\left(x_{ij},h\right),\\
\partial_{x_{i}}^{\alpha}W_{ij}=\partial_{x_{i}}^{\alpha}W\left(x_{ij},h\right).
\end{gather}
\end{subequations}The smoothing length $h$ connects the analytic
kernel in Eq. (\ref{eq:wendlandc4}) with the SPH kernels,
\begin{equation}
W\left(x-x^{\prime},h\right)=\psi\left(\frac{\sqrt{\left(x^{\alpha}-x^{\alpha,\prime}\right)\left(x^{\alpha,\prime}-x^{\alpha,\prime}\right)}}{h}\right).
\end{equation}

In practice, the smoothing length is not constant and instead has
discrete values at the SPH nodes. To retain symmetry in the derivatives
{[}such that $\partial_{x_{i}}^{\alpha}W\left(x_{i}-x_{j},h\right)=\partial_{x_{j}}^{\alpha}W\left(x_{j}-x_{i},h\right)${]},
the kernel values are taken to be an average of the evaluations using
these two smoothing lengths, \begin{subequations}

\begin{gather}
W_{ij}=\frac{1}{2}\left[W\left(x_{ij},h_{i}\right)+W\left(x_{ij},h_{j}\right)\right]\\
\partial_{x_{i}}^{\alpha}W_{ij}=\frac{1}{2}\left[\partial_{x_{i}}^{\alpha}W\left(x_{ij},h_{i}\right)+\partial_{x_{i}}^{\alpha}W\left(x_{ij},h_{j}\right)\right].
\end{gather}

The Wendland functions \cite{wendland1995piecewise} are one example
of a kernel that can be used in SPH. The Wendland C4 kernel is used
here,
\begin{equation}
W_{\text{wendland}}\left(x_{ij},h\right)=\begin{cases}
k\left(1-r\right)^{5}\left(8r^{2}+5r+1\right), & r\equiv x_{ij}/h\leq1,\\
0, & \text{otherwise},
\end{cases}\label{eq:wendlandc4}
\end{equation}
with a dimensionally-dependent normalization constant $k$ that is
chosen such that Eq. (\ref{eq:kernel-normalization}) is satisfied.
\end{subequations} \textcolor{black}{The choice of interpolation kernel is of course arbitrary for meshfree
methods such as these, and many have been used over the years. Wendland
kernels are common in many modern SPH studies (e.g. \cite{Read2010,Dehnen2012}),
but this choice is not necessarily unique.}

\textcolor{black}{Note in this section the per point volume $V_{j}$ has been used (such
as in Eq. \ref{eq:sph-interpolant}), leading to explicitly volume
weighted relations throughout. SPH does not define a volume element
per point as such, so in this paper the volume is defined as $V_{j}=m_{j}/\rho_{j}$.
It is common in the SPH literature to substitute this relation for
$V_{j}$ and then rearrange terms to remove the density from inside
the summation, such that only the fixed (and well defined) mass per
point remains (see for instance the discussion in Sec. 2.2 of \cite{monaghan2005smoothed}.)
In this work the volume remains inside the second derivative, which
may affect the accuracy of the method in regions of high density contrast.
This definition is consistent with the diffusion derivative in Refs.
\cite{brookshaw1985method} (where the mass of the particles is constant)
and \cite{monaghan2005smoothed} (where the mass is variable). It
is possible these results might be improved by rearranging such definitions
analogously to how first derivatives are typically handled, but such
an investigation is left to follow-on studies.}

\subsection{Spatial discretization}

To perform the spatial discretization, the material and radiation
energy equations {[}Eqs. (\ref{eq:ne-equations}){]} are multiplied
by $W\left(x_{i}-x^{\prime},h\right)$ and integrated,\begin{subequations}
\begin{gather}
\left\langle \frac{\rho}{\Delta t}\left(e^{n,\ell+1}-e^{n-1}\right)+c\sigma_{a}B^{n,\ell+1}-c\sigma_{a}E^{n,\ell+1}-Q_{e}^{n}\right\rangle _{i}=0,\\
\left\langle \frac{1}{\Delta t}E^{n,\ell+1}-\partial_{x}^{\alpha}\frac{c\lambda}{\sigma_{t}}\partial_{x}^{\alpha}E^{n,\ell+1}+c\sigma_{a}fE^{n,\ell+1}\right\rangle _{i}=\left\langle \frac{1}{\Delta t}E^{n-1}+c\sigma_{a}B^{n,\ell}-\left(1-f\right)c\sigma_{a}E^{n,\ell}+Q_{E}^{n}\right\rangle _{i},
\end{gather}
\end{subequations}where $\left\langle \cdot\right\rangle $ is the
interpolant notation as defined in Sec. \ref{subsec:sph-intro}. The
terms not involving derivatives can be approximated using Eq. (\ref{eq:interpolant-approx}),
e.g.
\begin{gather}
\left\langle \frac{\rho}{\Delta t}e^{n-1}\right\rangle \approx\frac{\rho_{i}}{\Delta t}e_{i}^{n-1},\\
\left\langle c\sigma_{a}fE^{n,\ell+1}\right\rangle _{i}\approx c\sigma_{a,i}f_{i}E_{i}^{n,\ell+1}.
\end{gather}
The second derivative in the radiation energy equation can be approximated
using Eq. (\ref{eq:sph-second-derivative}),
\begin{equation}
\left\langle -\partial_{x}^{\alpha}D\left(x\right)\partial_{x}^{\alpha}E\left(x\right)\right\rangle \approx-\sum_{j}V_{j}\left(D_{i}+D_{j}\right)\left(E_{i}-E_{j}\right)\frac{x_{ij}^{\alpha}}{x_{ij}^{\beta}x_{ij}^{\beta}}\partial_{x_{i}}^{\alpha}W_{ij},
\end{equation}
with the diffusion coefficient
\begin{equation}
D_{i}=\frac{c\lambda_{i}}{\sigma_{t,i}}.
\end{equation}
With these approximations, the material and radiation equations are\begin{subequations}\label{eq:fully-disc-mat-rad}

\begin{equation}
\frac{\rho_{i}}{\Delta t}\left(e_{i}^{n,\ell+1}-e_{i}^{n-1}\right)+c\sigma_{a,i}B_{i}^{n,\ell+1}-c\sigma_{a,i}E_{i}^{n,\ell+1}-Q_{e,i}^{n}=0,
\end{equation}
\begin{align}
\frac{1}{\Delta t}E_{i}^{n,\ell+1}-\sum_{j}V_{j}\left(D_{i}+D_{j}\right)\left(E_{i}^{n,\ell+1}-E_{j}^{n,\ell+1}\right)\frac{x_{ij}^{\alpha}}{x_{ij}^{\beta}x_{ij}^{\beta}}\partial_{x_{i}}^{\alpha}W_{ij}+c\sigma_{a,i}f_{i}E_{i}^{n,\ell+1}\nonumber \\
\qquad=\frac{1}{\Delta t}E_{i}^{n-1}+c\sigma_{a,i}B_{i}^{n,\ell}-\left(1-f_{i}\right)c\sigma_{a,i}E_{i}^{n,\ell}+Q_{E,i}^{n}.
\end{align}
 \end{subequations}The material energy equation is independent for
each point $i$, while the radiation energy equation for point $i$
linearly couples to the radiation energy from nearby nodes. For more
information on how these equations are solved, see Sec. \ref{subsec:mat-rad-solve}. 

\subsection{Conservation\label{subsec:conservation}}

The coupled material and radiation equations {[}Eqs. (\ref{eq:fully-disc-mat-rad}){]}
are conservative in energy to the tolerance of the nonlinear iteration
process once converged. The quantity that is conserved is the total
energy over all points, which is the material energy plus the radiation
energy,
\begin{equation}
\sum_{i}\left(m_{i}e_{i}+V_{i}E_{i}\right)=\text{const}.
\end{equation}
At convergence, $e^{n,\ell+1}\approx e^{n,\ell}$ and $E^{n,\ell+1}\approx E^{n,\ell}$,
which results in the simplified equations \begin{subequations}
\begin{gather}
\frac{\rho_{i}}{\Delta t}\left(e_{i}^{n}-e_{i}^{n-1}\right)+c\sigma_{a,i}B_{i}^{n}-c\sigma_{a,i}E_{i}^{n}=Q_{e,i}^{n},\\
\frac{1}{\Delta t}\left(E_{i}^{n}-E_{i}^{n-1}\right)-\sum_{j}V_{j}\left(D_{i}+D_{j}\right)\left(E_{i}^{n}-E_{j}^{n}\right)\frac{x_{ij}^{\alpha}}{x_{ij}^{\beta}x_{ij}^{\beta}}\partial_{x_{i}}^{\alpha}W_{ij}+c\sigma_{a,i}E_{i}^{n}-c\sigma_{a,i}B_{i}^{n}=Q_{E,i}^{n}.
\end{gather}
\end{subequations}Adding the two equations, multiplying by $V_{i}$,
and summing over $i$ results in 
\begin{equation}
\sum_{i}\left[m_{i}\left(e_{i}^{n}-e_{i}^{n-1}\right)+V_{i}\left(E_{i}^{n}-E_{i}^{n-1}\right)\right]=\Delta t\sum_{i}\left(Q_{e,i}^{n}+Q_{E,i}^{n}\right).
\end{equation}
Because the diffusion spatial derivative is antisymmetric about $i$
and $j$, it disappears under summation over both $i$ and $j$. This
final equation says that any gains or losses in the total energy are
as a result of specified sources. At equilibrium and absence external
sources, both the material and the radiation equations reduce to an
equilibrium absorption-emission rate at every point $x_{i}$, which
is $E_{i}=B_{i}$. 

\section{Methodology\label{sec:methodology}}

The thermal radiative transfer methods described here are implemented
within the open-source SPH code Spheral, described at \url{https://wci.llnl.gov/simulation/computer-codes/spheral}
and publicly available at \url{https://github.com/jmikeowen/spheral},
although the radiative transfer methods are not available in the open
source version. For access to the full code, including thermal radiative
transfer and input scripts for the problems presented in this paper,
please contact the authors. Spheral is written in C++ with a Python
interface, which allows for simple addition of new physics in either
programming language. The opacities, flux limiters, and equations
of state can be chosen arbitrarily for subsets of the SPH points as
needed, allowing problems with any number of distinct materials to
be studied.

\subsection{Meshless implementation}

In the succeeding examples Spheral's default method \cite{owen2010asph}
for computing the smoothing scale per point ($h_{i}$) is used with
the C4 Wendland kernel {[}Eq. (\ref{eq:wendlandc4}){]}. The target
radius of support is 4 points, i.e., the local smoothing scale for
each point is chosen to be 4 times the local particle spacing. \textcolor{black}{The smoothing scale algorithm from \cite{owen2010asph} is iterated
until the desired support for each point is achieved during problem
initialization. Since the points in these examples are not moving,
the smoothing scale is unchanged after this initialization. }

\textcolor{black}{Boundary conditions are handled with ghost nodes, which provide sufficient
support for the internal nodes and allow the boundary terms in SPH
to be neglected. The ghost points can be set to have constant values
independent of internal nodes or to represent an internal node. For
reflective or periodic boundaries, the points adjacent to the boundary
are copied across the boundary. The domain decomposition is done similarly,
where all neighboring points for those in the subdomain for a processor
are copied as ghost points. }

\subsection{Material and radiation energy solve\label{subsec:mat-rad-solve}}

The iterative process in Alg. \ref{alg:mat-rad-solve} \cite{brunner2020nonlinear}
can be written in terms of operator matrices as 
\begin{equation}
\left[\begin{array}{c}
e^{\ell+1}\\
E^{\ell+1}
\end{array}\right]=\left[\begin{array}{cc}
\mathbb{M}^{-1} & 0\\
0 & \mathcal{I}
\end{array}\right]\left[\begin{array}{cc}
\mathcal{R} & \mathcal{A}\\
0 & \mathcal{I}
\end{array}\right]\left[\begin{array}{cc}
\mathcal{I} & 0\\
0 & \mathcal{D}^{-1}
\end{array}\right]\left[\begin{array}{cc}
\mathcal{I} & 0\\
\mathbb{N} & \mathcal{S}
\end{array}\right]\left[\begin{array}{c}
e^{\ell}\\
E^{\ell}
\end{array}\right],\label{eq:matrix-iteration}
\end{equation}
with the operators defined by
\begin{lyxlist}{00.00.0000}
\begin{spacing}{0.8}
\item [{$\mathbb{M}^{-1}\left(P\right)$}] Solve the nonlinear material
equation, given a source $P_{i}$, 
\begin{equation}
\dfrac{\rho_{i}}{\Delta t}\left(e_{i}-e_{i}^{n-1}\right)+c\sigma_{a,i}aT_{i}^{4}=P_{i},\label{eq:material-energy-solve}
\end{equation}
 for each $e_{i}$,
\item [{$\mathcal{D}^{-1}\left(P\right)$}] Solve the diffusion equation,
given a source $P_{i}$, 
\begin{equation}
\left(\frac{1}{\Delta t}+c\sigma_{a,i}f_{i}\right)E_{i}-\sum_{j}V_{j}\left(D_{i}+D_{j}\right)\left(E_{i}-E_{j}\right)\frac{x_{ij}^{\alpha}}{x_{ij}^{\beta}x_{ij}^{\beta}}\partial_{x_{i}}^{\alpha}W_{ij}=P_{i},\label{eq:diffusion-solve}
\end{equation}
for all $E_{i}$,
\item [{$\mathcal{S}\left(E\right)$}] Calculate the radiation diffusion
source, $\dfrac{1}{\Delta t}E_{i}^{n-1}-\left(1-f_{i}\right)c\sigma_{a,i}E_{i}+Q_{E,i}$,
\item [{$\mathbb{N}\left(e\right)$}] Calculate the emission of radiation
from the material, $c\sigma_{a,i}aT_{i}^{4}$,
\item [{$\mathcal{A}\left(E\right)$}] Calculate the radiation energy absorption,
$c\sigma_{a,i}E_{i}$,
\item [{$\mathcal{R}\left(e\right)$}] Set the material energy source,
$Q_{e,i}^{n}$,
\end{spacing}
\end{lyxlist}
and the identity operator $\mathcal{I}$. This simplifies the addition
of new physics, reduces code duplication, and allows testing of each
operator independently. Note that $\mathbb{M}^{-1}$ and $\mathbb{N}$
are nonlinear operators, and should not be mistaken for matrices.
The operators $\mathcal{D}^{-1}$, $\mathcal{S},$and $\mathcal{A}$
are linear, but in practice, they are not explicitly formed into matrices.
The application of the combined operator in Eq. (\ref{eq:matrix-iteration})
is equivalent to a single outer iteration in Alg. \ref{alg:mat-rad-solve}.
The opacities and other material data ($\sigma_{a}$, $\sigma_{s}$,
$D$, $f$, $\lambda$, and $c_{v}$) are calculated at the start
of the time step and held constant within the time step. 

The material energy solve in Eq. (\ref{eq:material-energy-solve})
is done independently for each point using Newton's method, as shown
in Eq. (\ref{eq:full-material-solve}). Defining the function and
Jacobian for the material energy solve as\begin{subequations}\label{eq:material-energy-solve-operator}
\begin{gather}
u\left(e_{i}\right)=\frac{\rho}{\Delta t}\left(e_{i}-e^{n-1}\right)+c\sigma_{a}aT^{4}-P,\\
J_{m,e}\left(e\right)=\frac{\rho}{\Delta t}+\frac{4c\sigma_{a}aT^{3}}{c_{v}},
\end{gather}
the material energy is solved iteratively as
\begin{equation}
e_{i}^{k+1}=e_{i}^{k}-J_{m,e}^{-1}\left(e_{i}^{k}\right)u\left(e_{i}^{k}\right)\label{eq:material-energy-update}
\end{equation}
\end{subequations}until converged, where $k$ is the Newton iteration
index. As the call to an external equation of state can be expensive,
the temperature is updated for all points simultaneously, followed
by an update of the material energy for each point independently.
The iterative process in Eq. (\ref{eq:material-energy-update}) proceeds
until convergence each time the $\mathcal{M}^{-1}$ operator is called. 

The radiation energy solve in Eq. (\ref{eq:diffusion-solve}) is done
using the Hypre BoomerAMG preconditioner with GMRES \cite{falgout2002hypre}.
Because the material properties (opacities, specific heats, hydrodynamic
variables) are held constant within a time step, the preconditioner
can be initialized once at the start of each time step and then reused
each time the $\mathcal{D}^{-1}$ operator is called within the time
step. The radiation energy $E^{\ell}$ is used as an initial guess
for the GMRES solver when calculating $E^{\ell+1}$, which reduces
the number of GMRES iterations as the solution nears convergence,
as is discussed in Sec. \ref{subsec:manufactured-solution}. 

\subsection{Time step choice\label{subsec:time-step}}

Due to the implicit solve, the time step is updated not based on the
propagation speed of the radiation, but instead based on the observed
change in radiation energy over a time step. This is a heuristic for
the stability and accuracy of one Newton step. The idea is to limit
the fractional change in the material and radiation energies to some
finite value and increase or decrease the time step according to the
change from the previous time step. For example, for the material
energy, the fractional change in energy is calculated as 
\begin{equation}
\eta_{e}=\max_{i}\left(\frac{\left|e_{i}^{n}-e_{i}^{n-1}\right|}{e_{i}^{n}+\eta_{e}^{\text{target}}\bar{e}^{n}}\right),\label{eq:energy-fractional}
\end{equation}
where $\eta_{e}^{\text{target}}$ is the target fractional change
($\eta_{e}^{\text{target}}=0.05$ for the problems in this paper)
and $\bar{e}^{n}$ is the volume-averaged energy,
\begin{equation}
\bar{e}^{n}=\frac{\sum_{i}e_{i}^{n}V_{i}}{\sum_{i}V_{i}}.
\end{equation}
The $\bar{e}^{n}$ in the denominator of the fractional change prevents
points with very small radiation energy from dominating the time step
when changing a small magnitude relative to the other points in the
problem. Given the fractional change over the previous time step,
a proposed time step is calculated as
\begin{equation}
\Delta t_{e}^{n+1}=\Delta t^{n}\left(\frac{\eta_{e}^{\text{target}}}{\eta_{e}}\right)^{1/2}.
\end{equation}
\textcolor{black}{The fractional change in energy {[}Eq. (\ref{eq:energy-fractional}){]}
depends on the energy being nonzero somewhere in the problem. If it
is anticipated that the energy might be zero everywhere in the problem,
a small number (e.g. $10^{-15}$) can be added to the denominator
to prevent division by zero. }

The same is process is done for the radiation energy $E$, and the
time step is chosen to be the more restrictive of these, $\Delta t^{n+1}=\min\left(\Delta t_{e}^{n+1},\Delta t_{E}^{n+1}\right)$,
within the minimum and maximum time steps specified by the user. For
more information on time stepping strategies for thermal radiative
transfer with diffusion, including the one used here, see Ref. \cite{rider1999time}. 

\section{Results\label{sec:results}}

The results include three problems. The first is an infinite medium
equilibrium test, which is designed to show that the material and
radiation energies come to equilibrium at the correct rate, testing
the time discretization, emission, and absorption. The second is the
Su-Olson Marshak wave, which simulates the diffusion of a planar radiation
source into a vacuum, additionally testing the diffusion rate of the
radiation. The final problem is a manufactured solution that extends
the results to two and three dimensions. The tolerances used for each
of these problems are listed in Table \ref{tab:tolerances}, with
the inner tolerances referring to the convergence metric for ending
the Newton solve for the material energy and the GMRES solve for the
radiation energy and the outer tolerances used for ending the nonlinear
elimination iteration process (see Sec. \ref{subsec:mat-rad-solve}). 

\subsection{Infinite medium equilibrium test\label{subsec:infinite-medium}}

The first problem involves a single material in an infinite medium
in which the material and radiation energies are initially out of
equilibrium\textcolor{black}{, similar to the tests performed in Refs. \cite{whitehouse2004smoothed,whitehouse2005faster}}.
The units of length, time, temperature, and mass are chosen such that
the absorption opacity $\sigma_{a}$, speed of light $c$, black body
constant $a$, and ratio of proton mass to the Boltzmann constant,
$m_{p}/k_{B}$, are all one. The equation of state is chosen to be
an ideal gas \textcolor{black}{{[}Eq. (\ref{eq:ideal-eos}){]}}, which
in these units is defined as
\begin{equation}
T=\left(\gamma-1\right)\mu e.
\end{equation}
The problem can be written in the simplified unit system as a system
of two ordinary differential equations,
\begin{gather}
\rho\frac{\partial e}{\partial t}=E-\left(\alpha e\right)^{4},\\
\frac{\partial E}{\partial t}=-E+\left(\alpha e\right)^{4}.
\end{gather}
These equations are solved using a fifth-order Radau IIA integrator
\cite{hairer1999stiff} for comparison to the results given by the
code. 

The problem is run for two different cases, both of which have $\left(\gamma-1\right)\mu=1$
and a density of $\rho=1$, and in each case, the simulation proceeds
until $t_{end}=10$, which allows the material and radiation energies
to reach equilibrium. The first case starts with hot material and
cold radiation, while the second starts with cold material and hot
radiation, as shown in Table \ref{tab:infinite-initial}. The time
step is allowed to increase by up to an order of magnitude per time
step, up to the limit set by the changing material and radiation energies,
with a hard cap of $\Delta t_{max}=0.1$ (see Sec. \ref{subsec:time-step}). 

\begin{figure}
\centering{}\subfloat[Hot material at initial time]{\begin{centering}
\includegraphics[width=0.47\textwidth]{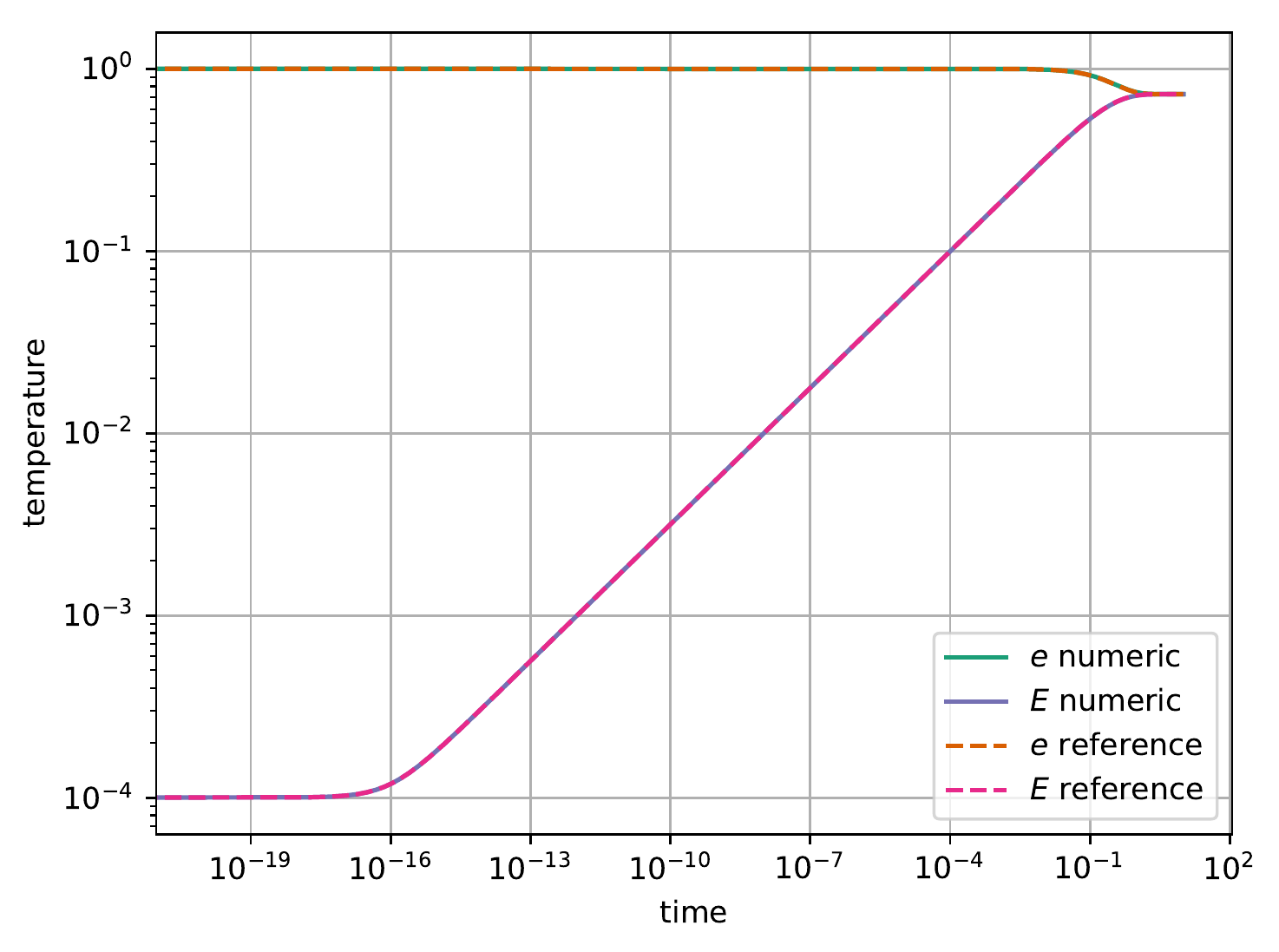}
\par\end{centering}
}\subfloat[Hot radiation at initial time]{\begin{centering}
\includegraphics[width=0.47\textwidth]{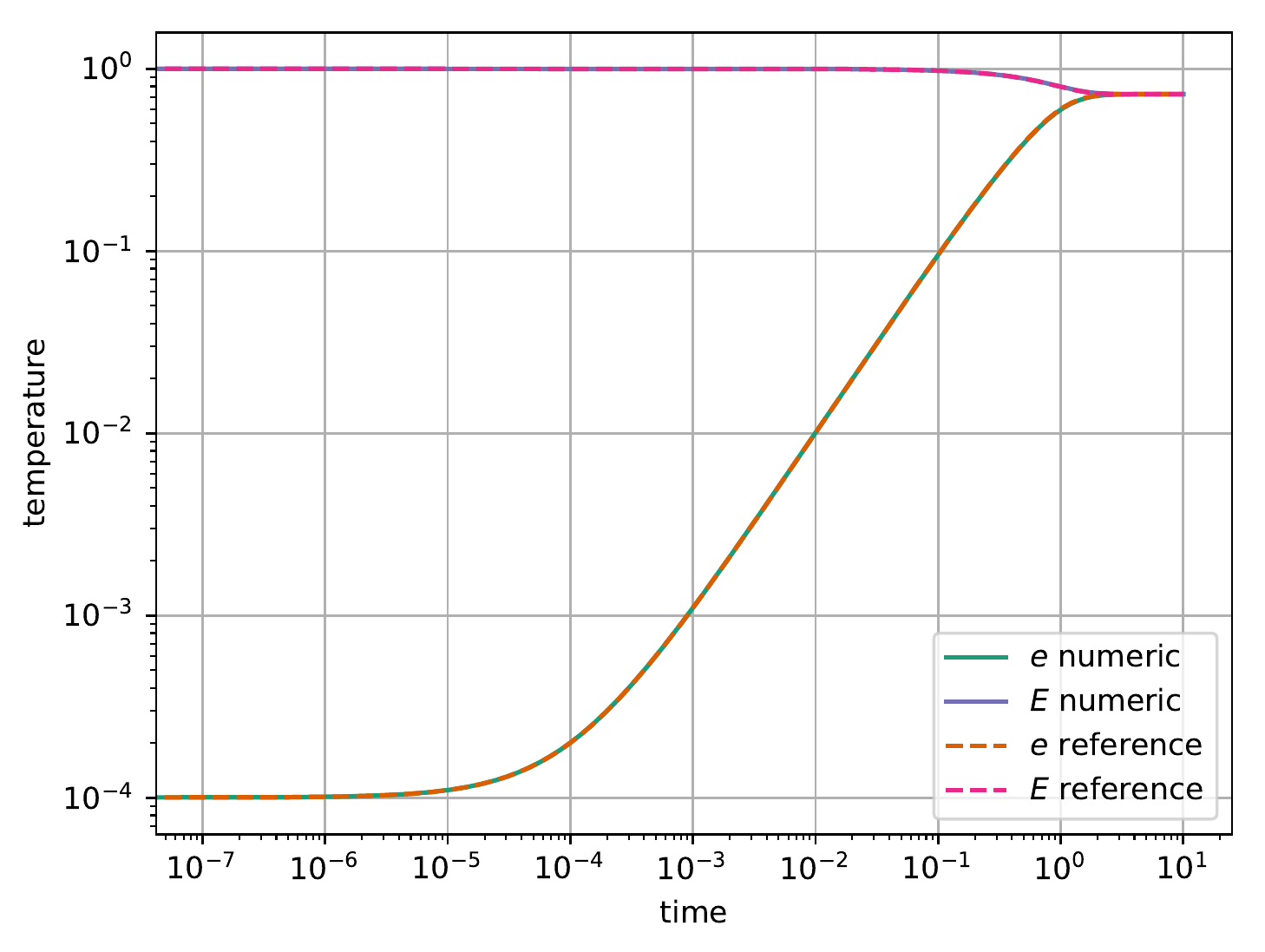}
\par\end{centering}
}\caption{Comparison of reference and numerical results for the infinite medium
problem. The time step is fixed at $\Delta t=0.0001$.}
\label{fig:infinite}
\end{figure}

\begin{figure}
\centering{}\includegraphics[width=0.47\textwidth]{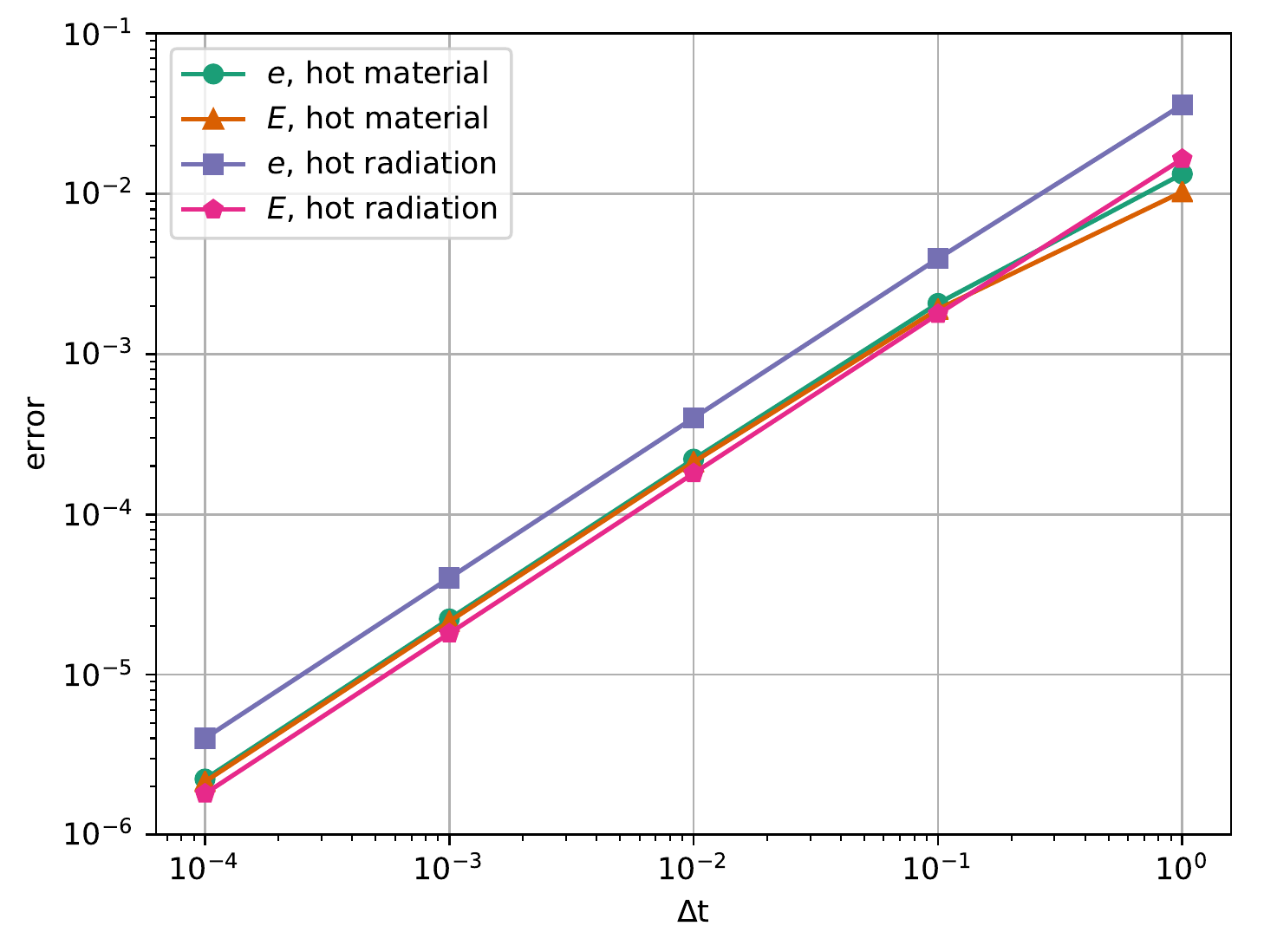}\caption{Convergence of the numerical solution for the infinite medium problem
to the reference solution with decreasing time step. The error is
calculated using Eq. (\ref{eq:infinite-error}).}
\label{fig:infinite-convergence}
\end{figure}

The first and second cases take, respectively, 779 and 155 time steps
to reach the goal time. The results are shown in Fig. \ref{fig:infinite}.
The reference and numeric solutions agree well for both problems.
The $L_{1}$ relative error for this problem is calculated with an
integral in time,
\begin{equation}
L_{1}\text{ error}=\frac{\int_{0}^{t_{end}}\left|T_{\text{ref}}-T_{\text{num}}\right|dt}{\int_{0}^{t_{end}}T_{\text{ref}}dt},\label{eq:infinite-error}
\end{equation}
using Simpson's rule evaluated at the time steps. The error between
the reference and numeric solutions for the first case (with hot material)
is $3.32\times10^{-4}$ for the material temperature and $2.55\times10^{-4}$
for the radiation temperature. The error calculated in the same way
for the second case (with hot radiation) is $1.22\times10^{-3}$ for
the material temperature and $6.21\times10^{-4}$ for the radiation
temperature. For a fixed time step $\Delta t$, the error decreases
linearly with $\Delta t$ as expected, as shown in Fig. \ref{fig:infinite-convergence}. 

The relative error between the initial and final energies ranges from
$1.11\times10^{-15}$ (hot radiation) and $4.44\times10^{-16}$ (hot
material) for $\Delta t=0.1$ to $4.22\times10^{-13}$ (hot radiation)
and $5.53\times10^{-13}$ (hot material) for $\Delta t=0.0001$, which
is of the correct order given the outer tolerance of $10^{-12}$ and
inner tolerance of $10^{-14}$ (Table \ref{tab:tolerances}).

\subsection{Su-Olson Marshak wave}

The second problem is a Marshak planar wave as described in Ref. \cite{su1997analytical}.
The problem consists of a reflecting plane at $x=0.0$ and a radiation
source from $0\leq x\leq x_{0}$ that is turned on for $0\leq\tau\leq\tau_{0}$,
where $x$ is a scaled distance and $\tau$ is a scaled time ($z$
is used in the paper for physical distance). The problem is started
with a material and radiation energy of $10^{-5}$ so the heat capacity
(an analytic $c_{v}\propto T^{3}$) is nonzero. \textcolor{black}{The equation of state from the original paper \cite{su1997analytical}
is 
\begin{equation}
e=\frac{4a}{\epsilon}T^{4},
\end{equation}
where $\epsilon$ is a time scaling factor. } The units for the problem are chosen such that $\sigma_{t}=1$, $\epsilon c=1$,
and $a=1$, which makes the scaled units referenced in the paper equal
to the physical units ($x=z$ and $\tau=t$). The solution to the
problem is semianalytic, and involves evaluating integrals to high
precision over $\eta\in\left[0,1\right]$ for each spatial point at
each time. For these results, due to the complicated structure of
the solution near $\eta=1$, a quadrature is first generated in $\xi\in\left[-1,1\right]$
and then transformed into $\eta$ via the relation
\begin{equation}
\eta=\frac{1}{\ell}\ln\left(\frac{1}{2}\left(1-\xi+e^{\ell}\left(1+\xi\right)\right)\right),
\end{equation}
in which the transformation becomes linear as $\ell\to0$ and concentrates
more points near $\eta=1$ as $\ell$ increases. A 5000-point Gauss-Legendre
quadrature with $\ell=3$ works well for the specific problem considered
below. The numerical result at the end time is compared to this semianalytic
result using an $L_{1}$ error,
\begin{equation}
L_{1}\text{ error}=\frac{\sum_{i}\left|e_{i}-e_{i}^{\text{semianalytic}}\right|}{\sum_{i}e_{i}^{\text{semianalytic}}},\label{eq:marshak-error}
\end{equation}
which differs from the standard approach of comparison to specific
$x$ and $\tau$ values tabulated in the paper. 

The parameters used for these results include an absorption fraction
of $c_{a}=0.5$, a time scaling factor of $\epsilon=1.0$, a radiation
source extent of $x_{0}=0.5$, and a scaled time at which the source
is turned off of $\tau_{0}=10.0$. The problem is run until $\tau=100.0$.
The spatial extent of the problem is $0\leq x\leq100$. 

\begin{figure}

\subfloat[Material energy]{\includegraphics[width=0.47\textwidth]{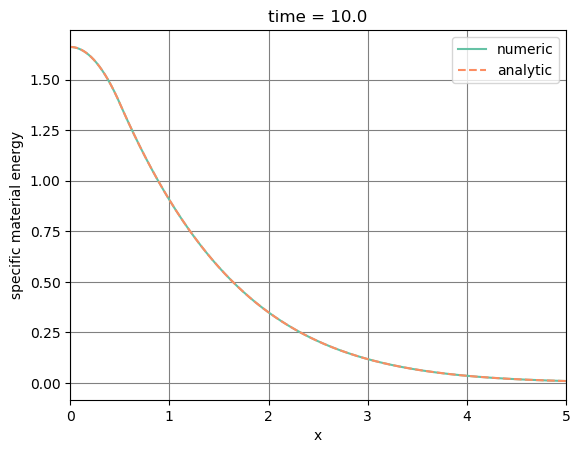}

}\subfloat[Radiation energy]{\includegraphics[width=0.47\textwidth]{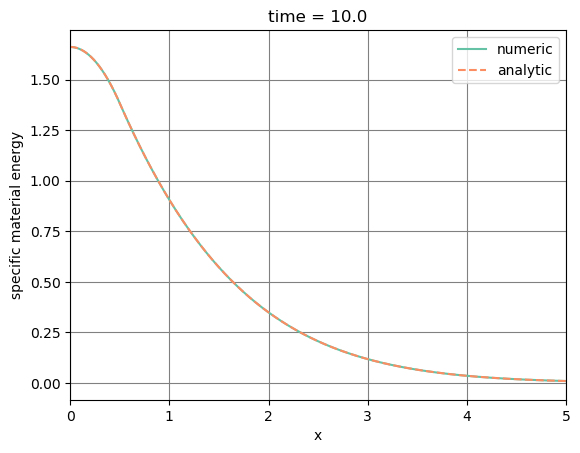}

}

\caption{The numeric and semianalytic solutions to the Marshak wave problem
at $t=10.0$. This figure is also available as a video.}
\label{fig:marshak-solution}
\end{figure}

\begin{figure}
\begin{centering}
\subfloat[Material energy]{\begin{centering}
\includegraphics[width=0.47\textwidth]{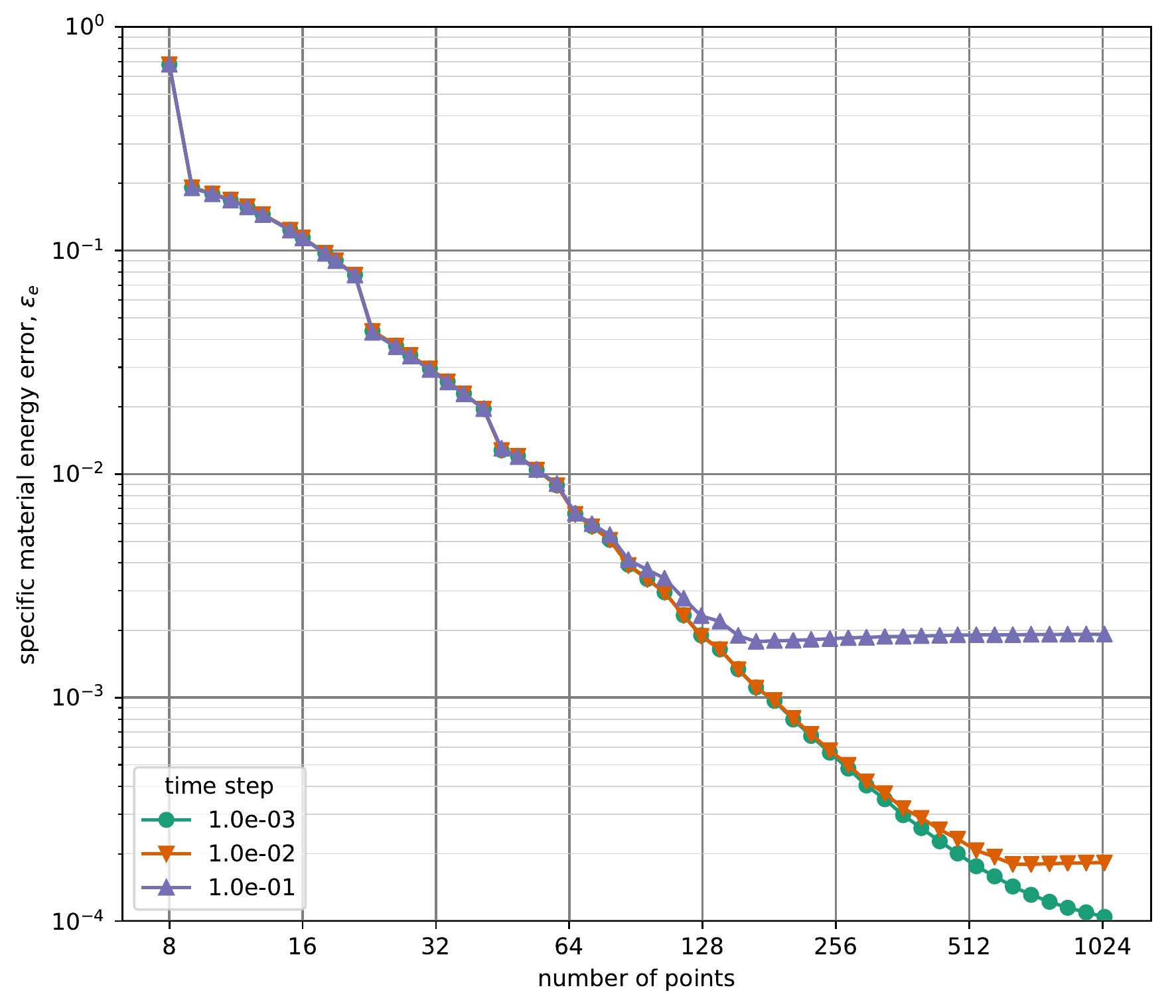}
\par\end{centering}
}\subfloat[Radiation energy]{\begin{centering}
\includegraphics[width=0.47\textwidth]{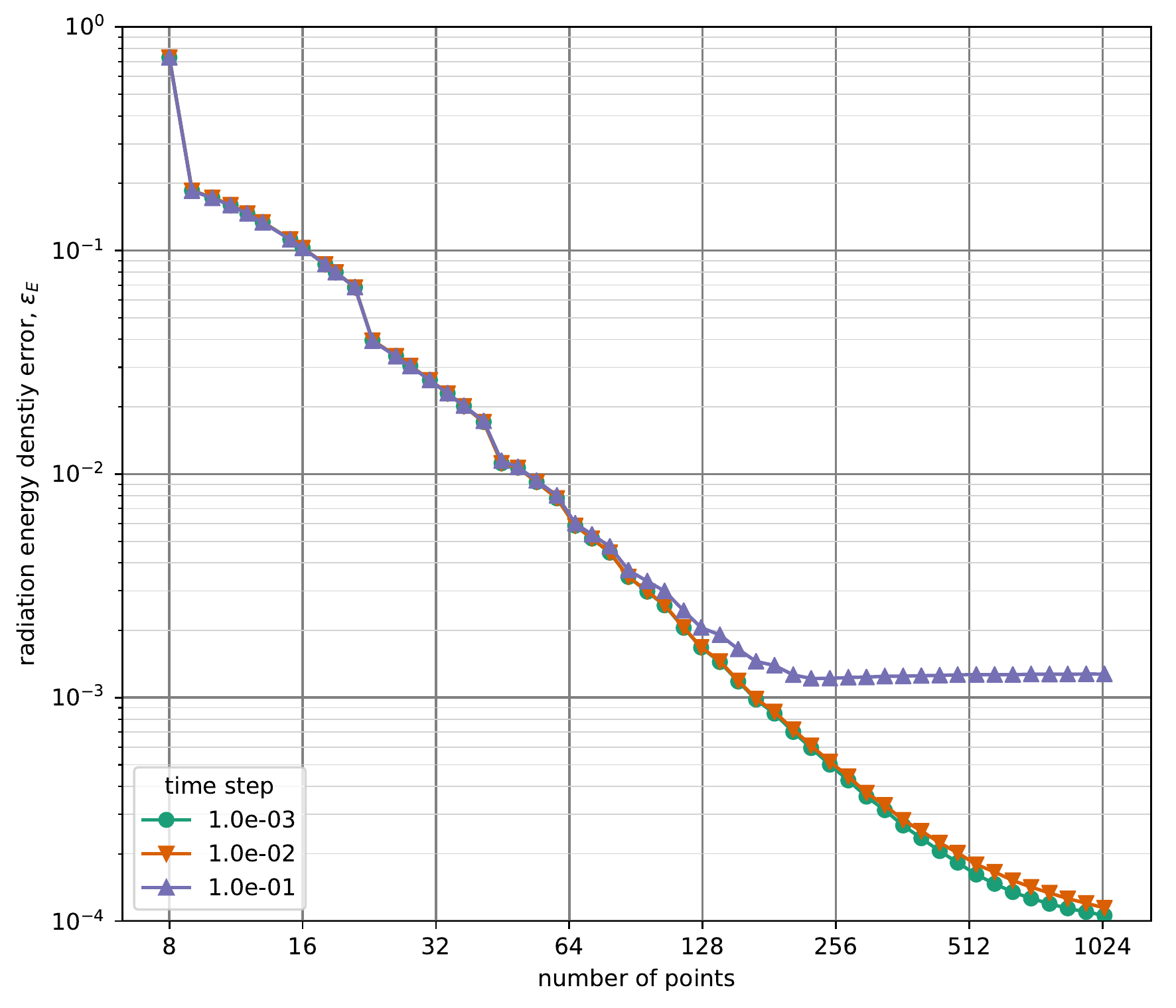}
\par\end{centering}
}
\par\end{centering}
\centering{}\caption{Convergence of the numeric solution to the semianalytic solution for
the Marshak wave with an increasing number of spatial points for three
time step values. The error is calculated using Eq. (\ref{eq:marshak-error}).}
\label{fig:marshak}
\end{figure}

The time evolution of the semianalytic and numeric solutions is shown
in Fig. \ref{fig:marshak-solution}. The spatial convergence results
are shown in Fig. \ref{fig:marshak} for three different time steps.
The code converges to the semianalytic result with second-order accuracy
spatially as the point spacing decreases and with first-order accuracy
temporally as the time step decreases, as expected based on SPH spatial
interpolation and backward Euler time integration.

\subsection{Manufactured solution for material-radiation coupling\label{subsec:manufactured-solution}}

This problem tests the material-radiation coupling using the method
of manufactured solutions (MMS). To apply MMS to Eqs. (\ref{eq:fully-disc-mat-rad}),
solutions for $e$ and $E$ are chosen for the equations, \begin{subequations}\label{eq:manufactured-solutions}
\begin{gather}
e_{m}=e_{0}\left[1.2+\prod_{\alpha=1}^{\text{dim}}\cos\left(2\pi\frac{x_{\alpha}-v_{m}t}{d_{m}}\right)\right],\\
E_{m}=E_{0}\left[1.2+\prod_{\alpha=1}^{\text{dim}}\cos\left(2\pi\frac{x_{\alpha}+v_{m}t}{d_{m}}-\omega_{m}\right)\right],
\end{gather}
\end{subequations}with $e_{0}=10^{13}\text{ ergs\ensuremath{\cdot}g}^{-1}$,
$E_{0}=7a\left(e_{0}/c_{v}\right)^{4}\text{ ergs\ensuremath{\cdot}cm}^{-3}$,
$v_{m}=5\times10^{9}\text{ cm\ensuremath{\cdot}s}^{-1}$, $d_{m}=5\text{ cm}$,
and $\omega_{m}=1/16$. The opacities are set to constant values of
$\sigma_{a}=0.05\text{ cm}^{-1}$ and $\sigma_{s}=0.95\text{ cm}^{-1}$.
These solutions are inserted into the non-discretized equations {[}Eqs.
(\ref{eq:material-energy}) and (\ref{eq:diffusion}){]} to calculate
sources, $Q_{e}$ and $Q_{E}$. The solutions are chosen such that
the magnitude of the individual terms in Eqs. (\ref{eq:material-energy})
and (\ref{eq:diffusion}) is approximately equal, with the opposite
time movement of the waves $\pm v$ allowing each individual piece
of physics to dominate at certain times. The spatial extent of the
problem is set to one full wavelength, or $d_{m}$ in each dimension
(to allow for periodic boundary conditions), and the problem is run
for one cycle, or until $t_{end}=d_{m}/v_{m}$, using a time step
of $\Delta t=t_{end}/1000$. At $t_{end}$, the numerical solution
is compared to the analytic solution in Eqs. (\ref{eq:manufactured-solutions})
to calculate the $L_{1}$ relative error,
\begin{equation}
L_{1}\text{ error}=\frac{\sum_{i}\left|e_{i}-e_{m,i}\right|}{\sum_{i}e_{m,i}}\label{eq:manufactured-error}
\end{equation}
(with a similar equation for $E$). \textcolor{black}{The ideal gas equation of state is used {[}Eq. (\ref{eq:ideal-eos}){]}
with $\gamma=5/3$ and $\alpha=1$.}

\begin{figure}

\subfloat[Material energy]{\includegraphics[width=0.47\textwidth,trim={2cm 4cm 7cm 2cm},clip]{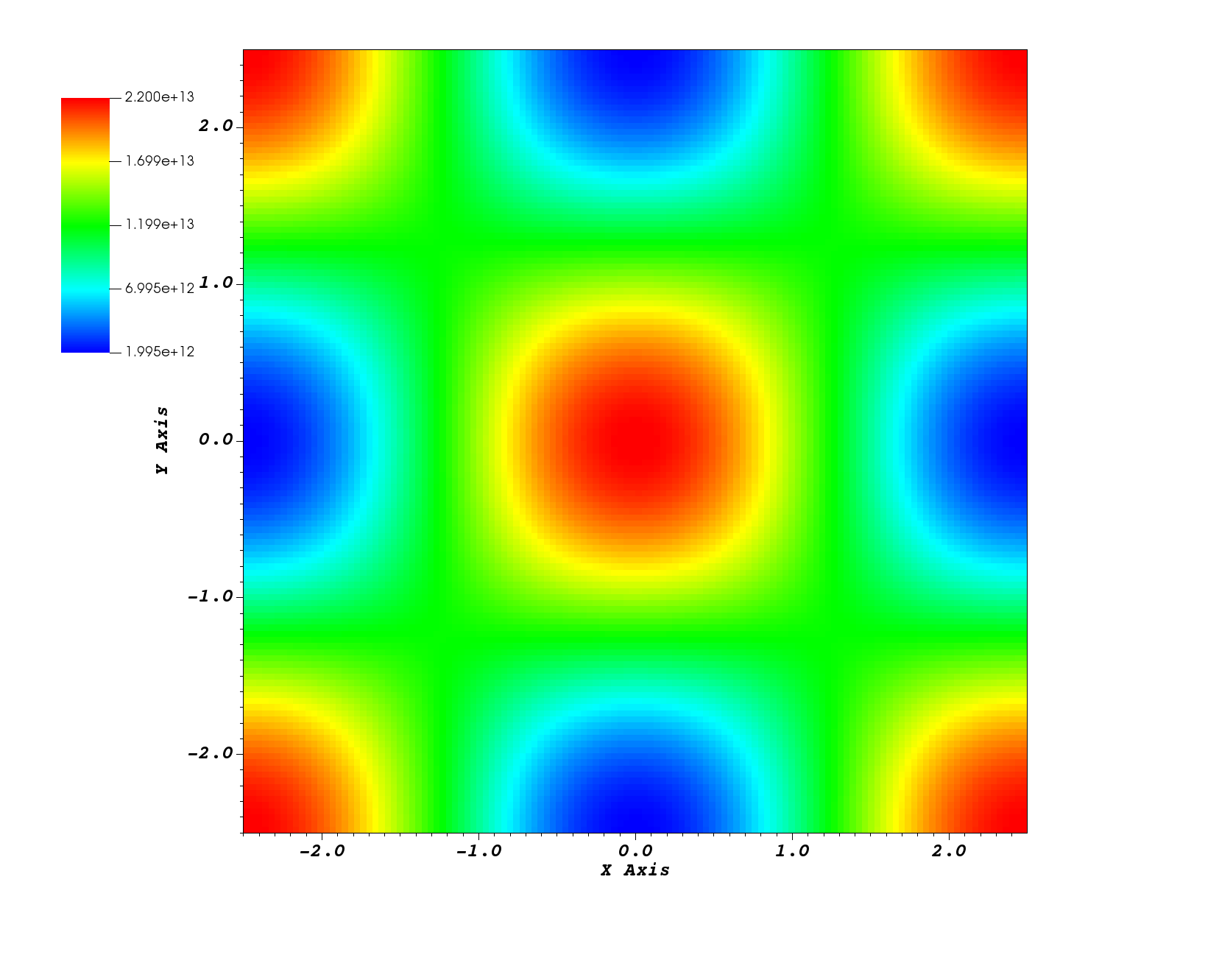}

}\subfloat[Radiation energy]{\includegraphics[width=0.47\textwidth,trim={2cm 4cm 7cm 2cm},clip]{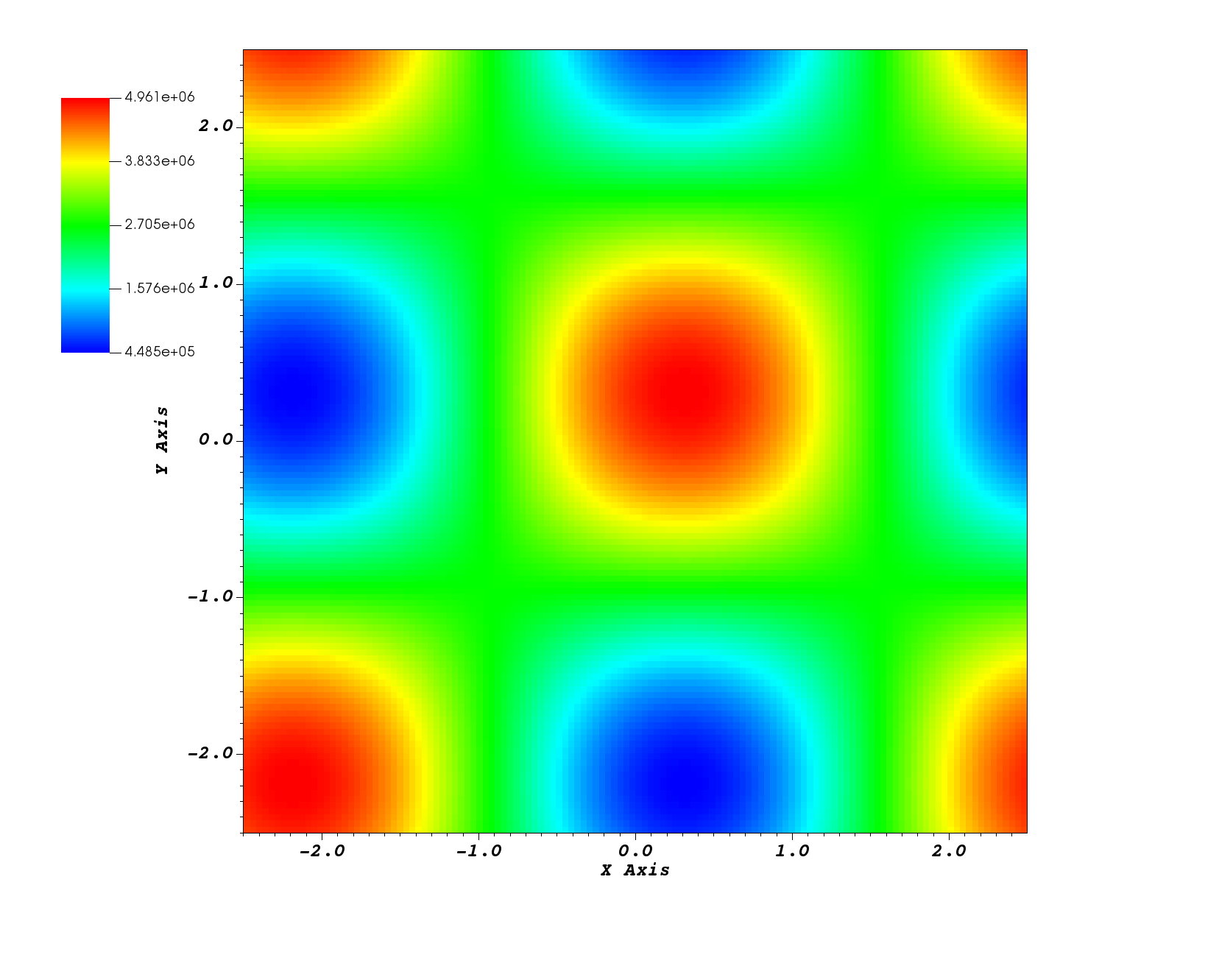}

}

\caption{The numeric solution to the manufactured problem in 2D for $128^{2}$
points after one cycle. This figure is also available as a video.}
\label{fig:manufactured-solution-2d}
\end{figure}

\begin{figure}
\begin{centering}
\subfloat[Material energy]{\begin{centering}
\includegraphics[width=0.47\textwidth]{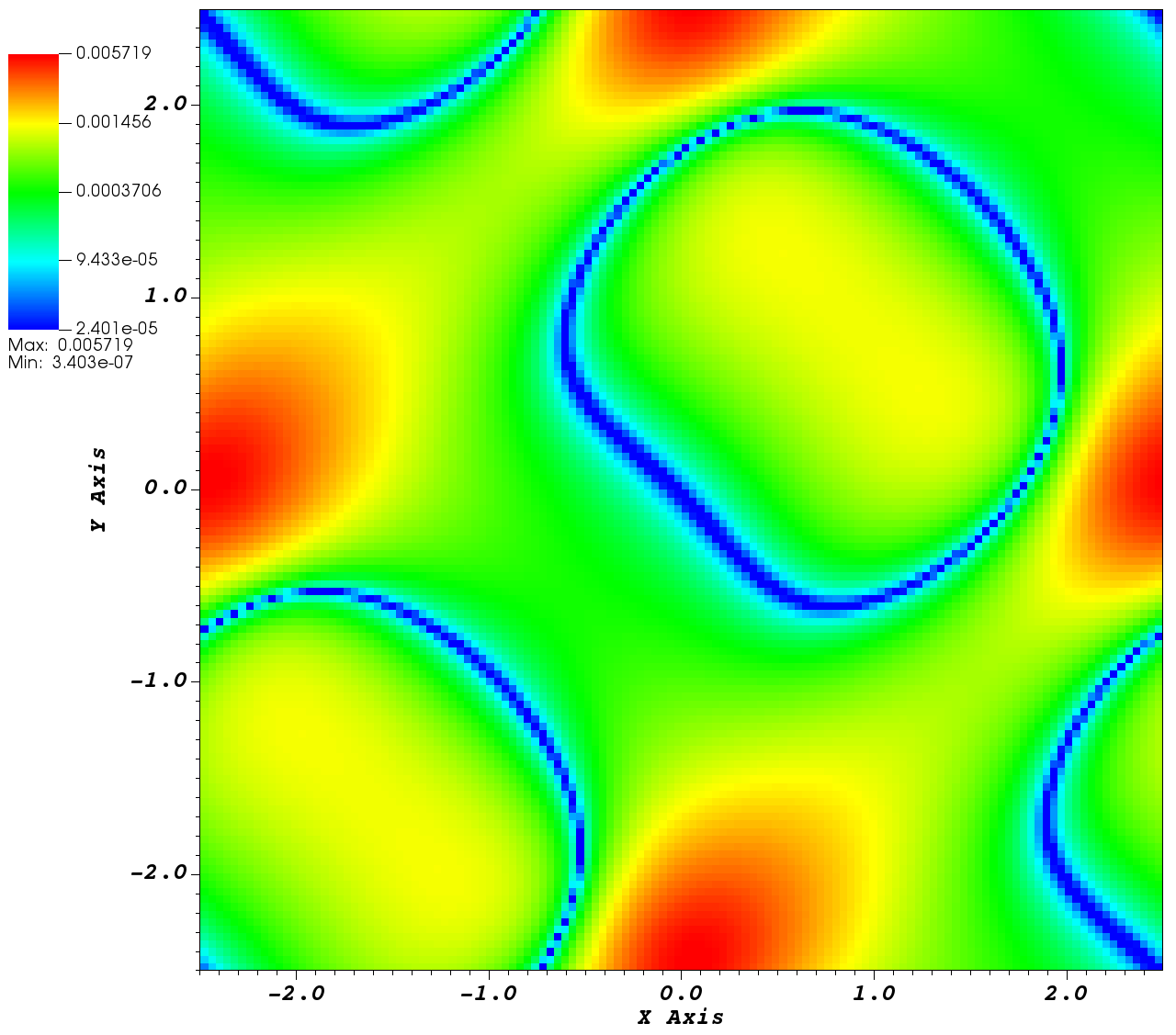}
\par\end{centering}
}\subfloat[Radiation energy]{\begin{centering}
\includegraphics[width=0.47\textwidth]{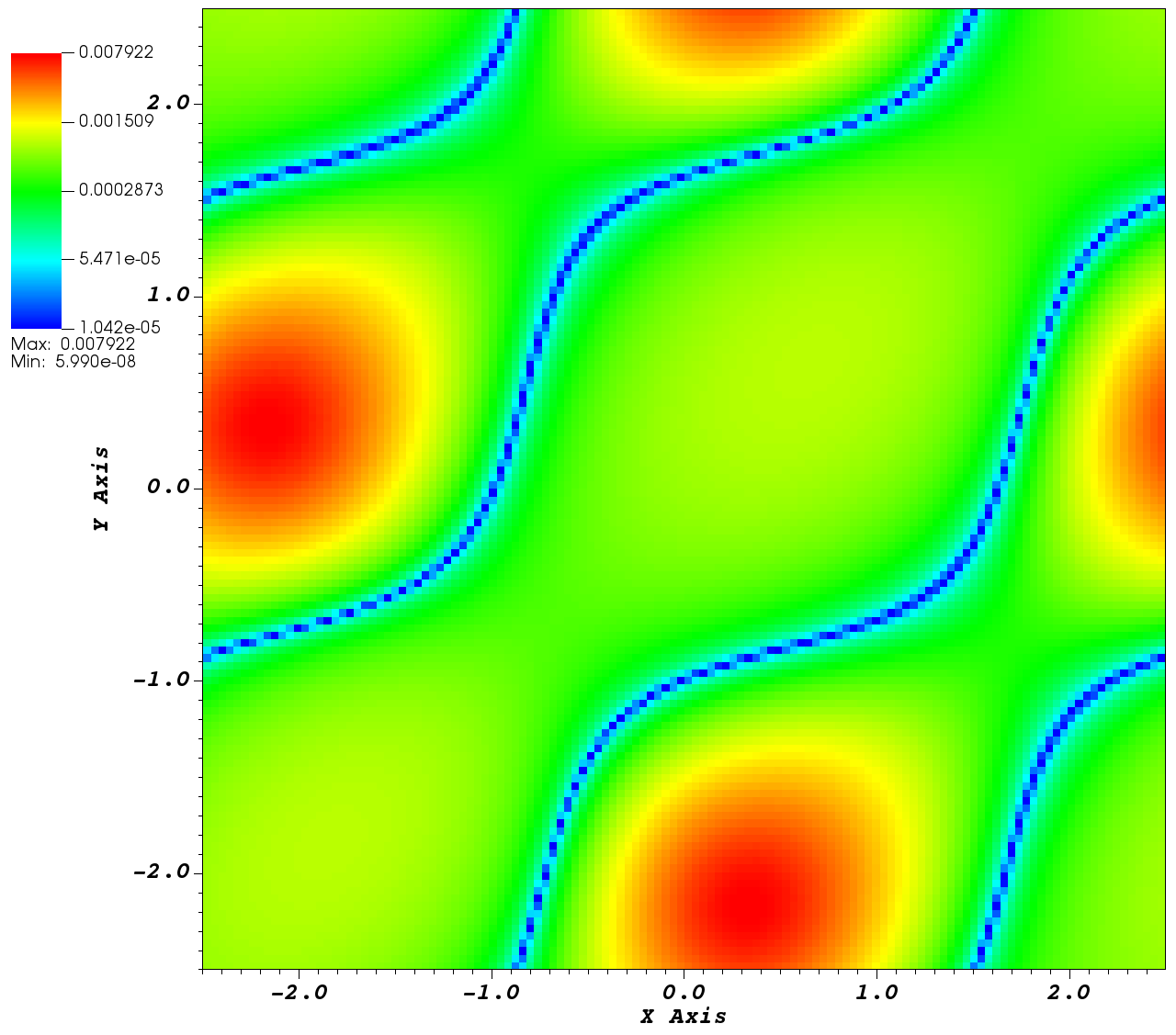}
\par\end{centering}
}
\par\end{centering}
\centering{}\caption{Pointwise relative error of the numeric solution to the manufactured
problem at $10^{-9}$ s in 2D for $128^{2}$ points.}
\label{fig:manufactured-error}
\end{figure}

The solution for the problem with $128^{2}$ points is shown in Fig.
\ref{fig:manufactured-solution-2d}. At the end time, the solution
should be equal to the initial condition. The relative error between
the solution after one cycle (at $10^{-9}$ s) and the manufactured
solution for the specific thermal energy and radiation energy density
is shown in Fig. \ref{fig:manufactured-error}. The relative error
is highest where the solution is lowest, with maxima of 0.006 for
the specific material energy and 0.008 for the radiation energy. For
much of the domain, the relative error is below 0.001 for both the
material and radiation energy. At the end time, the relative difference
between the starting and ending energy of the system is $1.67\times10^{-12}$,
which indicates good energy conservation.

\begin{figure}
\begin{centering}
\subfloat[Material energy]{\begin{centering}
\includegraphics[width=0.47\textwidth]{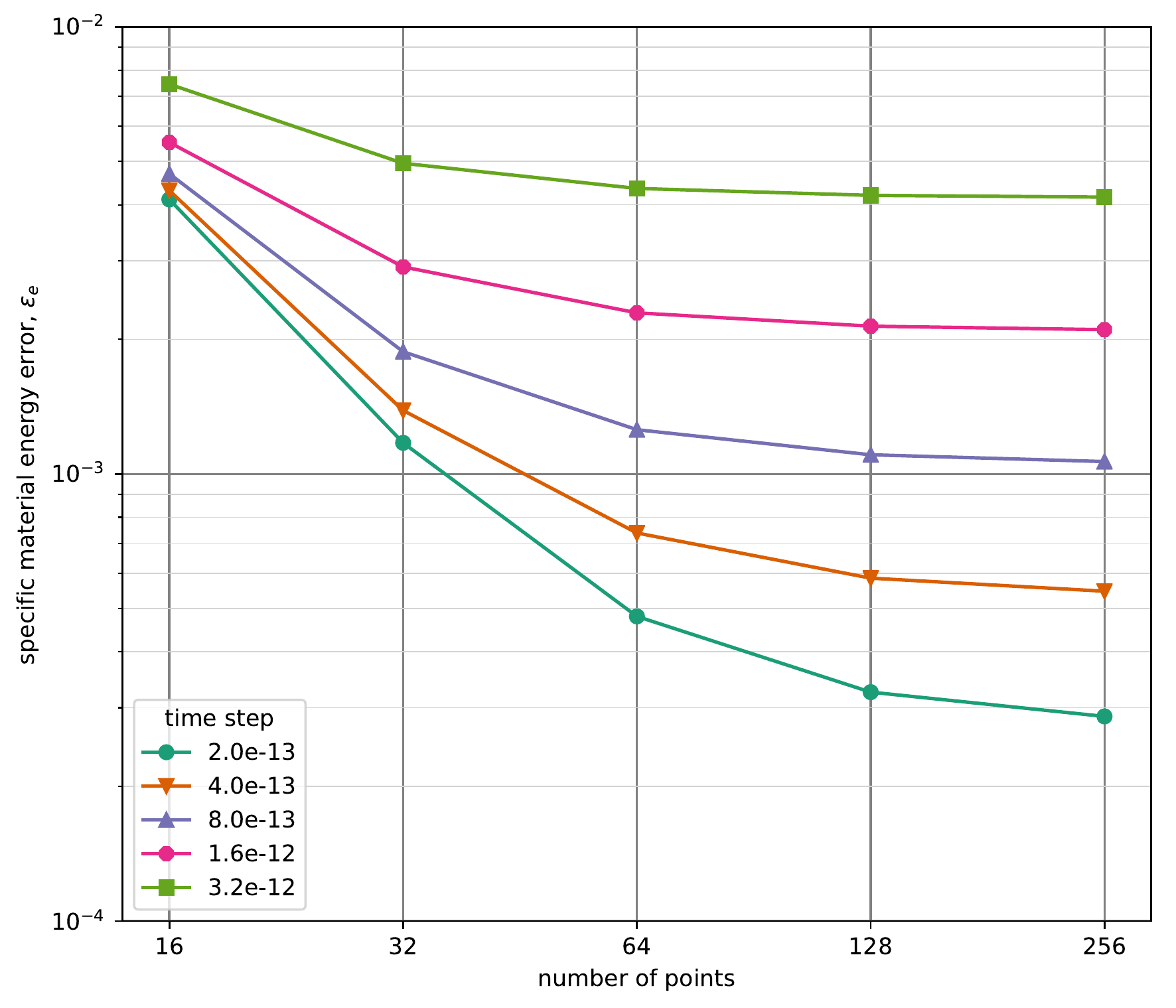}
\par\end{centering}
}\subfloat[Radiation energy]{\begin{centering}
\includegraphics[width=0.47\textwidth]{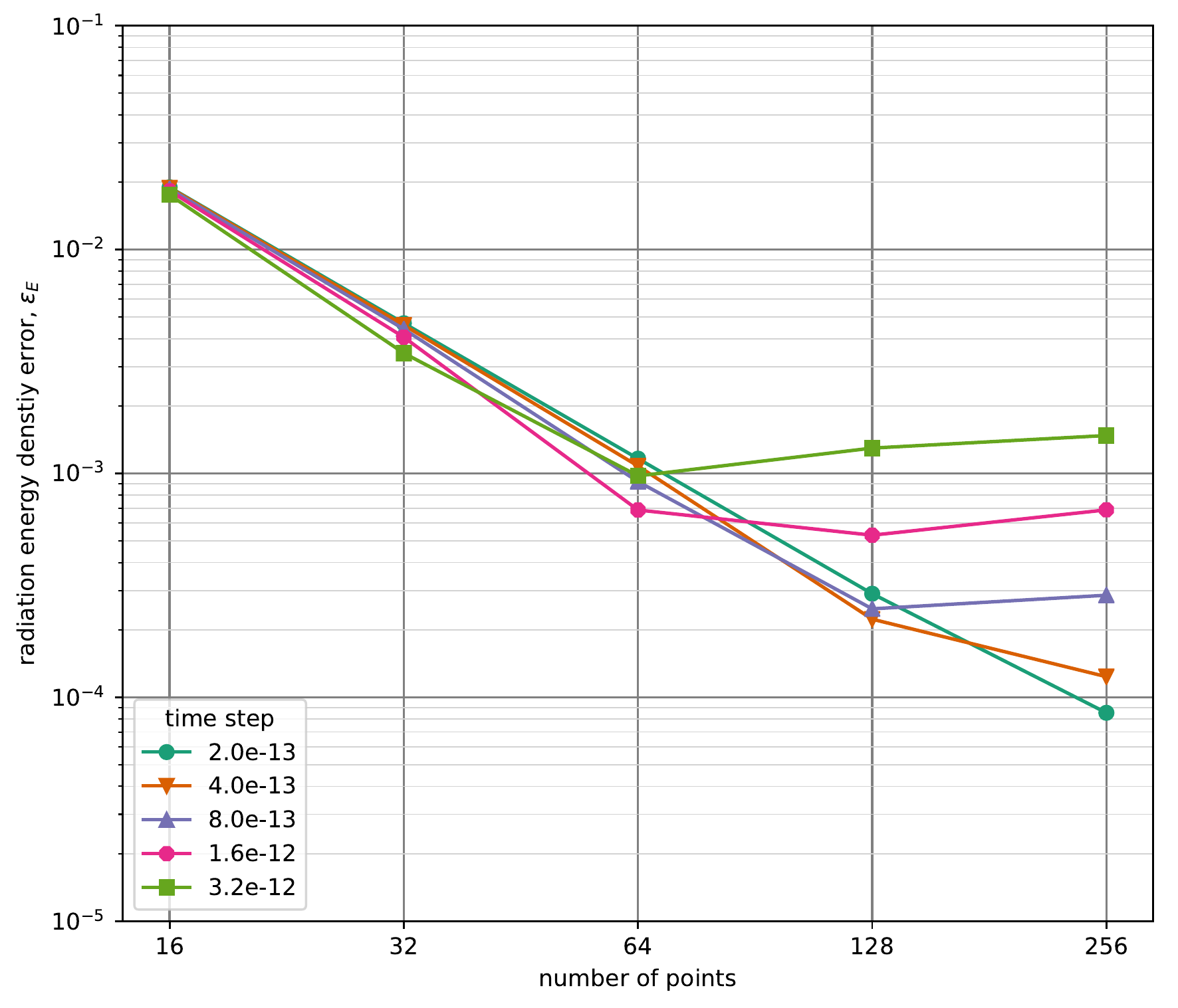}
\par\end{centering}
}
\par\end{centering}
\centering{}\caption{Convergence of the numeric solution to the manufactured problem for
several time step values in 1D. The error is calculated using Eq.
(\ref{eq:manufactured-error}). \protect\textcolor{black}{Note that the material energy is more dependent on the time step for
convergence than the radiation energy, and thus does not show the
same second-order convergence for the time steps shown here. }}
\label{fig:manufactured-convergence-1d}
\end{figure}

The $L_{1}$ error for the manufactured problem in 1D for several
time step values is shown in Fig. \ref{fig:manufactured-convergence-1d}.
The numeric solution converges to the manufactured solution with second-order
accuracy in space and first-order accuracy in time (again as expected).
For a time step of $2\times10^{-13}$, the error in the radiation
energy indicates second-order convergence for all values of point
spacing. The error in the material energy, which only has spatial
coupling through the radiation, is much more dependent on the time
step for this problem than the radiation energy. As in the previous
problems, the error decreases linearly with the time step, when not
limited by the spatial error. 

\begin{figure}
\begin{centering}
\subfloat[Material energy]{\begin{centering}
\includegraphics[width=0.47\textwidth]{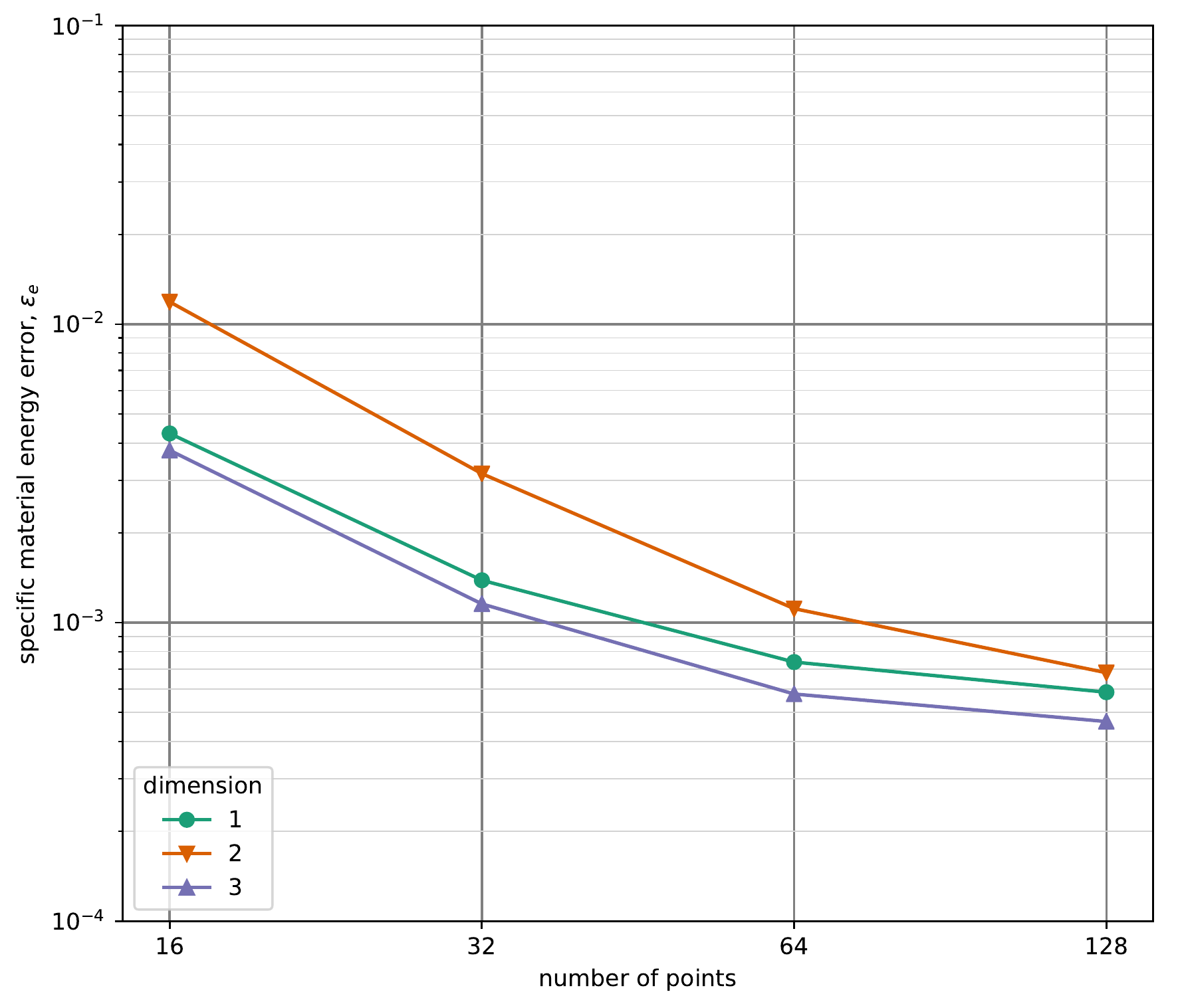}
\par\end{centering}
}\subfloat[Radiation energy]{\begin{centering}
\includegraphics[width=0.47\textwidth]{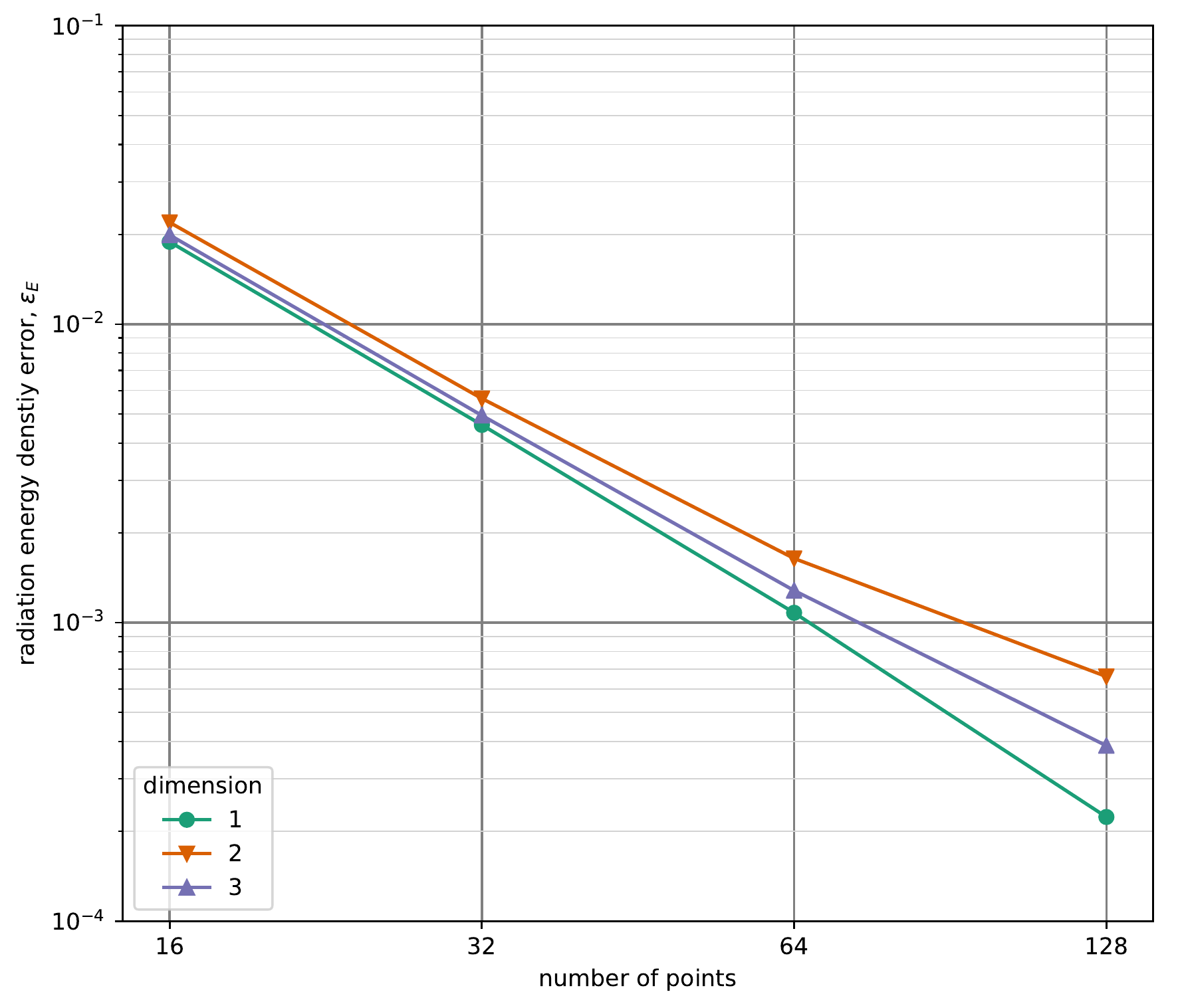}
\par\end{centering}
}
\par\end{centering}
\centering{}\caption{Convergence of the numeric solution to the manufactured problem in
1D, 2D, and 3D with a fixed time step of $\Delta t=4\times10^{-13}$.
The error is calculated using Eq. (\ref{eq:manufactured-error}).}
\label{fig:manufactured-convergence}
\end{figure}

Based on the results in 1D, a time step of $4\times10^{-13}$ is chosen
for the comparison of 1D, 2D, and 3D results up to $128^{d}$ points
(for the dimension $d$). The convergence results for 1D, 2D, and
3D are shown in Fig. \ref{fig:manufactured-convergence}. The 2D and
3D results have similar errors compared to the manufactured solution
for similar spacing of points, although not identical due to differences
in the interpolation functions for 2D and 3D, and also show second-order
convergence when not limited by the time step. 

While the number of processors used for each case was not chosen to
show scaling with a constant number of points per processor or a constant
number of total points, the performance results in Table \ref{tab:manufactured-timing}
show some general scaling trends. In 2D, the small number of points
per processor means that communication costs dominate, and adding
more points per each processor helps reduce that cost. In 3D, the
cases with 910 points per processor show around a 74 percent efficiency
when the number of points and processors are quadrupled. With many
more points per processor than the other cases, the 2,097,152-point
problem on 576 processors runs with a similar efficiency to the 262,144-point
problem in 288 processors. \textcolor{black}{Note that the 3D results take a long time to run due to the high level
of support, up to several hundred neighbors per point.}

The number of outer iterations per time step is governed for this
problem by the tolerance. For the tolerance given in Table \ref{tab:tolerances},
the solver needed two outer iterations to converge (and a third to
check for convergence) in each time step. The number of GMRES iterations
per outer iteration is around 3-4 on average. At the start of a time
step, when the guess for the material and radiation energies is based
on the values from the previous time step, the number is generally
higher, around 6-8. On the last outer iteration, the number of GMRES
solves is 1, indicating that the guess from the previous iteration
satisfies the diffusion equation. 

The solution to the problem is the same at the start and end times,
so despite adding and subtracting energy in certain regions of the
problem, the initial and final energies should be equal. The relative
error between the initial and final energy in 2D ranges from $6.35\times10^{-14}$
for the 256 points to $1.69\times10^{-12}$ for 16,384 points. For
3D, the relative error ranges from $1.33\times10^{-13}$ for 4,096
points to $4.53\times10^{-12}$ for 2,097,152 points. In both cases,
the relative error is on the order of the tolerance for the radiation
solve of $10^{-12}$ and much lower than the tolerance of the coupled
radiation and material energy solve of $10^{-8}$. 

\section{Conclusions and future work\label{sec:conclusions}}

The thermal radiative transfer equations with grey diffusion is discretized
fully implicitly in time and solved using an efficient nonlinear elimination
method that leads to fast convergence of the emission source. A standard
SPH diffusion derivative is applied to the modified equations to form
a fully discretized set of equations that are conservative in energy.
The diffusion term is solved using optimized linear solvers, which
permits good parallel efficiency. 

The code is verified by comparison to three test problems with known
solutions. The code gets the correct infinite medium behavior for
either a hot material emitting radiation or a cold material absorbing
radiation, with the expected first-order convergence in time. The
results for a one-dimensional semianalytic Marshak wave problem are
consistent with second-order spatial convergence. Finally, the code
is run in 2D and 3D for a manufactured problem with sinusoidal radiation
and material energy solutions traveling in opposite directions, which
likewise exhibits second-order convergence in space and first-order
convergence in time. 

The diffusion discretization for SPH is stable and performs well,
but lacks zeroth-order consistency. Other methods, such as reproducing
kernel particle methods \cite{liu1995reproducing} and moving least
squares particle hydrodynamics \cite{dilts1999moving,dilts2000moving},
do not lack the zeroth-order consistency and perform better near boundaries.
Conservative reproducing kernel smoothed particle hydrodynamics is
implemented in the SPH code used for these results \cite{frontiere2017crksph},
and a similar diffusion discretization would make the hydrodynamics
and radiation discretizations consistent. Other discretizations would
also make the application of vacuum or incoming radiation boundary
conditions more feasible. 

\section*{Acknowledgements}

This work was performed under the auspices of the U.S. Department
of Energy by Lawrence Livermore National Laboratory under Contract
DE-AC52-07NA27344. This document was prepared as an account of work
sponsored by an agency of the United States government. Neither the
United States government nor Lawrence Livermore National Security,
LLC, nor any of their employees makes any warranty, expressed or implied,
or assumes any legal liability or responsibility for the accuracy,
completeness, or usefulness of any information, apparatus, product,
or process disclosed, or represents that its use would not infringe
privately owned rights. Reference herein to any specific commercial
product, process, or service by trade name, trademark, manufacturer,
or otherwise does not necessarily constitute or imply its endorsement,
recommendation, or favoring by the United States government or Lawrence
Livermore National Security, LLC. The views and opinions of authors
expressed herein do not necessarily state or reflect those of the
United States government or Lawrence Livermore National Security,
LLC, and shall not be used for advertising or product endorsement
purposes. LLNL-JRNL-799713. 

\bibliographystyle{unsrt}
\bibliography{sph_rad_hydro_refs}

\begin{thebibliography}{10}

\bibitem{monaghan2005smoothed}
Joe~J. Monaghan.
\newblock Smoothed particle hydrodynamics.
\newblock {\em Reports on progress in physics}, 68(8):1703, 2005.

\bibitem{liu2010smoothed}
MB~Liu and GR~Liu.
\newblock Smoothed particle hydrodynamics ({SPH}): an overview and recent
  developments.
\newblock {\em Archives of computational methods in engineering}, 17(1):25--76,
  2010.

\bibitem{morris1996study}
Joseph~Peter Morris.
\newblock A study of the stability properties of smooth particle hydrodynamics.
\newblock {\em Publications of the Astronomical Society of Australia},
  13(1):97--102, 1996.

\bibitem{dilts1999moving}
Gary~A Dilts.
\newblock Moving-least-squares-particle hydrodynamics {I}: Consistency and
  stability.
\newblock {\em International Journal for Numerical Methods in Engineering},
  44(8):1115--1155, 1999.

\bibitem{dilts2000moving}
Gary~A Dilts.
\newblock Moving least-squares particle hydrodynamics {II}: conservation and
  boundaries.
\newblock {\em International Journal for Numerical Methods in Engineering},
  48(10):1503--1524, 2000.

\bibitem{frontiere2017crksph}
Nicholas Frontiere, Cody~D Raskin, and J~Michael Owen.
\newblock {CRKSPH}--a conservative reproducing kernel smoothed particle
  hydrodynamics scheme.
\newblock {\em Journal of Computational Physics}, 332:160--209, 2017.

\bibitem{whitehouse2004smoothed}
Stuart~C Whitehouse and Matthew~R Bate.
\newblock Smoothed particle hydrodynamics with radiative transfer in the
  flux-limited diffusion approximation.
\newblock {\em Monthly Notices of the Royal Astronomical Society},
  353(4):1078--1094, 2004.

\bibitem{whitehouse2005faster}
Stuart~C Whitehouse, Matthew~R Bate, and Joe~J Monaghan.
\newblock A faster algorithm for smoothed particle hydrodynamics with radiative
  transfer in the flux-limited diffusion approximation.
\newblock {\em Monthly Notices of the Royal Astronomical Society},
  364(4):1367--1377, 2005.

\bibitem{bate2015combining}
Matthew~R Bate and Eric~R Keto.
\newblock Combining radiative transfer and diffuse interstellar medium physics
  to model star formation.
\newblock {\em Monthly Notices of the Royal Astronomical Society},
  449(3):2643--2667, 2015.

\bibitem{viau2006implicit}
Serge Viau, Pierre Bastien, and Seung-Hoon Cha.
\newblock An implicit method for radiative transfer with the diffusion
  approximation in smooth particle hydrodynamics.
\newblock {\em The Astrophysical Journal}, 639(1):559, 2006.

\bibitem{mayer2007fragmentation}
Lucio Mayer, Graeme Lufkin, Thomas Quinn, and James Wadsley.
\newblock Fragmentation of gravitationally unstable gaseous protoplanetary
  disks with radiative transfer.
\newblock {\em The Astrophysical Journal Letters}, 661(1):L77, 2007.

\bibitem{brookshaw1985method}
Leigh Brookshaw.
\newblock A method of calculating radiative heat diffusion in particle
  simulations.
\newblock {\em Publications of the Astronomical Society of Australia},
  6(2):207--210, 1985.

\bibitem{petkova2009implementation}
Margarita Petkova and Volker Springel.
\newblock An implementation of radiative transfer in the cosmological
  simulation code gadget.
\newblock {\em Monthly Notices of the Royal Astronomical Society},
  396(3):1383--1403, 2009.

\bibitem{altay2008sphray}
Gabriel Altay, Rupert~AC Croft, and Inti Pelupessy.
\newblock Sphray: a smoothed particle hydrodynamics ray tracer for radiative
  transfer.
\newblock {\em Monthly Notices of the Royal Astronomical Society},
  386(4):1931--1946, 2008.

\bibitem{nayakshin2009dynamic}
Sergei Nayakshin, Seung-Hoon Cha, and Alexander Hobbs.
\newblock Dynamic monte carlo radiation transfer in {SPH}: radiation pressure
  force implementation.
\newblock {\em Monthly Notices of the Royal Astronomical Society},
  397(3):1314--1325, 2009.

\bibitem{fryer2006snsph}
Christopher~L Fryer, Gabriel Rockefeller, and Michael~S Warren.
\newblock Snsph: a parallel three-dimensional smoothed particle radiation
  hydrodynamics code.
\newblock {\em The Astrophysical Journal}, 643(1):292, 2006.

\bibitem{lowrie2004comparison}
Robert~B Lowrie.
\newblock A comparison of implicit time integration methods for nonlinear
  relaxation and diffusion.
\newblock {\em Journal of Computational Physics}, 196(2):566--590, 2004.

\bibitem{mousseau2000physics}
VA~Mousseau, DA~Knoll, and WJ~Rider.
\newblock Physics-based preconditioning and the newton--krylov method for
  non-equilibrium radiation diffusion.
\newblock {\em Journal of computational physics}, 160(2):743--765, 2000.

\bibitem{morel1985synthetic}
Jim~E Morel, Edward~W Larsen, and MK~Matzen.
\newblock A synthetic acceleration scheme for radiative diffusion calculations.
\newblock {\em Journal of Quantitative Spectroscopy and Radiative Transfer},
  34(3):243--261, 1985.

\bibitem{morel2007linear}
Jim~E Morel, T-Y~Brian Yang, and James~S Warsa.
\newblock Linear multifrequency-grey acceleration recast for preconditioned
  krylov iterations.
\newblock {\em Journal of Computational Physics}, 227(1):244--263, 2007.

\bibitem{brown2001preconditioning}
Peter~N Brown and Carol~S Woodward.
\newblock Preconditioning strategies for fully implicit radiation diffusion
  with material-energy transfer.
\newblock {\em SIAM Journal on Scientific Computing}, 23(2):499--516, 2001.

\bibitem{tetsu2016comparison}
Hiroyuki Tetsu and Taishi Nakamoto.
\newblock Comparison of implicit schemes to solve equations of radiation
  hydrodynamics with a flux-limited diffusion approximation: Newton--raphson,
  operator splitting, and linearization.
\newblock {\em The Astrophysical Journal Supplement Series}, 223(1):14, 2016.

\bibitem{lanzkron1996analysis}
Paul~J Lanzkron, Donald~J Rose, and James~T Wilkes.
\newblock An analysis of approximate nonlinear elimination.
\newblock {\em SIAM Journal on Scientific Computing}, 17(2):538--559, 1996.

\bibitem{brunner2020nonlinear}
Thomas~A Brunner, Terry~S Haut, and Paul~F Nowak.
\newblock Nonlinear elimination applied to radiation diffusion.
\newblock {\em Nuclear Science and Engineering}, pages 1--13, 2020.

\bibitem{sadat2006use}
Hamou Sadat.
\newblock On the use of a meshless method for solving radiative transfer with
  the discrete ordinates formulations.
\newblock {\em Journal of Quantitative Spectroscopy and Radiative Transfer},
  101(2):263--268, 2006.

\bibitem{liu2007meshless}
LH~Liu and JY~Tan.
\newblock Meshless local {P}etrov-{G}alerkin approach for coupled radiative and
  conductive heat transfer.
\newblock {\em International journal of thermal sciences}, 46(7):672--681,
  2007.

\bibitem{kindelan2010application}
Manuel Kindelan, Francisco Bernal, Pedro Gonz{\'a}lez-Rodr{\'\i}guez, and
  Miguel Moscoso.
\newblock Application of the {RBF} meshless method to the solution of the
  radiative transport equation.
\newblock {\em Journal of Computational Physics}, 229(5):1897--1908, 2010.

\bibitem{bassett2019meshless}
Brody Bassett and Brian Kiedrowski.
\newblock Meshless local {P}etrov--{G}alerkin solution of the neutron transport
  equation with streamline-upwind petrov--galerkin stabilization.
\newblock {\em Journal of Computational Physics}, 377:1--59, 2019.

\bibitem{rokrok2012element}
B~Rokrok, H~Minuchehr, and A~Zolfaghari.
\newblock Element-free {G}alerkin modeling of neutron diffusion equation in
  x--y geometry.
\newblock {\em Annals of Nuclear Energy}, 43:39--48, 2012.

\bibitem{tanbay2013numerical}
Tayfun Tanbay and Bilge Ozgener.
\newblock Numerical solution of the multigroup neutron diffusion equation by
  the meshless {RBF} collocation method.
\newblock {\em Mathematical and Computational Applications}, 18(3):399--407,
  2013.

\bibitem{birdsall1991particle}
Charles~K Birdsall.
\newblock Particle-in-cell charged-particle simulations, plus monte carlo
  collisions with neutral atoms, pic-mcc.
\newblock {\em IEEE Transactions on plasma science}, 19(2):65--85, 1991.

\bibitem{falgout2002hypre}
Robert~D Falgout and Ulrike~Meier Yang.
\newblock hypre: A library of high performance preconditioners.
\newblock In {\em International Conference on Computational Science}, pages
  632--641. Springer, 2002.

\bibitem{levermore1984relating}
CD~Levermore.
\newblock Relating eddington factors to flux limiters.
\newblock {\em Journal of Quantitative Spectroscopy and Radiative Transfer},
  31(2):149--160, 1984.

\bibitem{morel2000diffusion}
JE~Morel.
\newblock Diffusion-limit asymptotics of the transport equation, the {P}1/3
  equations, and two flux-limited diffusion theories.
\newblock {\em Journal of Quantitative Spectroscopy and Radiative Transfer},
  65(5):769--778, 2000.

\bibitem{mihalas2013foundations}
Dimitri Mihalas and Barbara~Weibel Mihalas.
\newblock {\em Foundations of radiation hydrodynamics}.
\newblock Courier Corporation, 2013.

\bibitem{fleckjr1971implicit}
JA~Fleck~Jr and JD~Cummings~Jr.
\newblock An implicit monte carlo scheme for calculating time and frequency
  dependent nonlinear radiation transport.
\newblock {\em Journal of Computational Physics}, 8(3):313--342, 1971.

\bibitem{wendland1995piecewise}
Holger Wendland.
\newblock Piecewise polynomial, positive definite and compactly supported
  radial functions of minimal degree.
\newblock {\em Advances in computational Mathematics}, 4(1):389--396, 1995.

\bibitem{Read2010}
J~I Read, T~Hayfield, and O~Agertz.
\newblock {Resolving mixing in smoothed particle hydrodynamics}.
\newblock {\em Monthly Notices of the Royal Astronomical Society},
  405(3):1513--1530, jul 2010.

\bibitem{Dehnen2012}
Walter Dehnen and Hossam Aly.
\newblock {Improving convergence in smoothed particle hydrodynamics simulations
  without pairing instability}.
\newblock {\em Monthly Notices of the Royal Astronomical Society},
  425(2):1068--1082, sep 2012.

\bibitem{owen2010asph}
J~Michael Owen.
\newblock {ASPH} modeling of material damage and failure.
\newblock In {\em Proceedings of the 5th International {SPHERIC} Workshop},
  pages 297--304, Manchester, UK, Jan 2010.

\bibitem{rider1999time}
William~J Rider and Dana~A Knoll.
\newblock Time step size selection for radiation diffusion calculations.
\newblock {\em Journal of Computational Physics}, 152(2):790--795, 1999.

\bibitem{hairer1999stiff}
Ernst Hairer and Gerhard Wanner.
\newblock Stiff differential equations solved by radau methods.
\newblock {\em Journal of Computational and Applied Mathematics},
  111(1-2):93--111, 1999.

\bibitem{su1997analytical}
Bingjing Su and Gordon~L Olson.
\newblock An analytical benchmark for non-equilibrium radiative transfer in an
  isotropically scattering medium.
\newblock {\em Annals of Nuclear Energy}, 24(13):1035--1055, 1997.

\bibitem{liu1995reproducing}
Wing~Kam Liu, Sukky Jun, and Yi~Fei Zhang.
\newblock Reproducing kernel particle methods.
\newblock {\em International journal for numerical methods in fluids},
  20(8-9):1081--1106, 1995.

\end{thebibliography}

\pagebreak{}

\begin{algorithm}
\begin{algorithmic}[1]
\State{$\ell=0$}
\State{set initial guesses to $e^{n,\ell}=e^{n-1}$ and $E^{n,\ell}=E^{n-1}$}
\State{solve $m\left(e^{n,\ell},E^{n,\ell}\right)=0$ for $e^{n,\ell}$}
\State{\quad (perform Newton updates until converged to inner tolerance for $e$)}
\While{$e^n$ and $E^n$ not converged to outer tolerances}
\State{solve $r^{\dagger}\left(e^{n,\ell},E^{n,\ell},E^{n,\ell+1}\right)=0$ for $E^{n,\ell+1}$}
\State{\quad (if iterative solver is used, iterate until converged to inner tolerance for $E$)}
\State{solve $m\left(e^{n,\ell+1},E^{n,\ell+1}\right)=0$ for $e^{n,\ell+1}$}
\State{\quad (perform Newton updates until converged to inner tolerance for $e$)}
\State{$\ell=\ell+1$}
\EndWhile
\end{algorithmic}

\caption{Material and radiation energy update, which is based on Ref. \cite{brunner2020nonlinear}.
The material and radiation energy equations $m$ and $r^{\dagger}$
are defined in Eq. (\ref{eq:ne-equations}), while the Newton update
for $m$ is described in Eq. (\ref{eq:full-material-solve}). The
convergence criteria are defined in Eqs. (\ref{eq:mat-rad-convergence}). }
\label{alg:mat-rad-solve}
\end{algorithm}

\begin{table}
\begin{centering}
\begin{tabular}{|c|c|c|c|c|}
\hline 
\multirow{2}{*}{Problem} & \multicolumn{2}{c|}{Inner} & \multicolumn{2}{c|}{Outer}\tabularnewline
\cline{2-5} \cline{3-5} \cline{4-5} \cline{5-5} 
 & Material & Radiation & Material & Radiation\tabularnewline
\hline 
\hline 
Infinite medium & $10^{-14}$ & $10^{-14}$ & $10^{-12}$ & $10^{-12}$\tabularnewline
\hline 
Marshak wave & $10^{-12}$ & $10^{-12}$ & $10^{-8}$ & $10^{-8}$\tabularnewline
\hline 
Manufactured problem & $10^{-12}$ & $10^{-12}$ & $10^{-8}$ & $10^{-8}$\tabularnewline
\hline 
\end{tabular}
\par\end{centering}
\centering{}\caption{Convergence tolerances for the results. Then inner tolerance is for
the independent radiation and material energy solves, while the outer
tolerance is for the combined solve.}
\label{tab:tolerances}
\end{table}

\begin{table}
\begin{centering}
\begin{tabular}{|c|c|c|c|c|c|}
\hline 
Case & $e_{0}$ & $E_{0}$ & $T_{e,0}$ & $T_{r,0}$ & $\Delta t_{init}$\tabularnewline
\hline 
\hline 
Hot material & $1.0$ & $10^{-16}$ & $1.0$ & $10^{-4}$ & $10^{-20}$\tabularnewline
\hline 
Hot radiation & $10^{-4}$ & $1.0$ & $10^{-4}$ & $1.0$ & $10^{-7}$\tabularnewline
\hline 
\end{tabular}
\par\end{centering}
\centering{}\caption{Initial conditions for infinite medium problem for two cases, one
starting with a high thermal temperature and low radiation temperature
and one with the temperatures reversed. See Sec. \ref{subsec:infinite-medium}
for units.}
\label{tab:infinite-initial}
\end{table}

\begin{table}
\begin{tabular}{|c|c|c|c|c|c|c|c|}
\hline 
Dimension & Points & Procs & Points/proc & GMRES/Outer & Outer/Step & Wall time (s) & Time$\times$Proc/Point\tabularnewline
\hline 
\hline 
\multirow{4}{*}{2} & 256 & 36 & 7 & 3.66 & 3 & 59 & 8.29\tabularnewline
\cline{2-8} \cline{3-8} \cline{4-8} \cline{5-8} \cline{6-8} \cline{7-8} \cline{8-8} 
 & 1,024 & 36 & 28 & 4.66 & 3 & 79 & 2.78\tabularnewline
\cline{2-8} \cline{3-8} \cline{4-8} \cline{5-8} \cline{6-8} \cline{7-8} \cline{8-8} 
 & 4,096 & 36 & 113 & 3.33 & 3 & 261 & 2.29\tabularnewline
\cline{2-8} \cline{3-8} \cline{4-8} \cline{5-8} \cline{6-8} \cline{7-8} \cline{8-8} 
 & 16,384 & 36 & 455 & 3.66 & 3 & 817 & 1.79\tabularnewline
\hline 
\multirow{4}{*}{3} & 4096 & 36 & 113 & 3.66 & 3 & 815 & 7.21\tabularnewline
\cline{2-8} \cline{3-8} \cline{4-8} \cline{5-8} \cline{6-8} \cline{7-8} \cline{8-8} 
 & 32,768 & 36 & 910 & 4.66 & 3 & 5,610 & 6.20\tabularnewline
\cline{2-8} \cline{3-8} \cline{4-8} \cline{5-8} \cline{6-8} \cline{7-8} \cline{8-8} 
 & 262,144 & 288 & 910 & 3.33 & 3 & 7,660 & 8.41\tabularnewline
\cline{2-8} \cline{3-8} \cline{4-8} \cline{5-8} \cline{6-8} \cline{7-8} \cline{8-8} 
 & 2,097,152 & 576 & 3,640 & 3.66 & 3 & 31,975 & 8.78\tabularnewline
\hline 
\end{tabular}

\caption{Timing for the manufactured problem in 3D.}
\label{tab:manufactured-timing}
\end{table}

\end{document}